\documentclass[12pt]{article}
\usepackage{amssymb}
\usepackage{epsfig}

\advance\textheight by \topskip
\newcommand{\text}[1]{\mbox{#1}}
\input{tcilatex}

\begin{document}

\title{CMB-slow, or How to Estimate Cosmological\\
Parameters by Hand}
\author{V. Mukhanov \\
LMU, Sektion Physik, Theresienstr.37, 80333 Muenchen }
\maketitle

\begin{abstract}
I derive analytically the spectrum of the CMB fluctuations. The final result
for $C_{l}$ is presented in terms of elementary functions with an explicit
dependence on the basic cosmological parameters. This result is in a rather
good agreement with CMBFAST for a wide range of parameters around
concordance model. This allows us to understand the physical reasons for
dependence of the particular features of the CMB spectrum on the basic
cosmological parameters and to estimate the possible accuracy of their
determination. I also analyse the degeneracy of the spectrum with respect to
certain combinations of the cosmological parameters.
\end{abstract}

\section{\protect\bigskip Introduction}

After recombination the primordial radiation doesn't interact anymore with
the matter and most of the photons come to us without further scattering.
Since the radiation is extremely isotropic in nearly all angular scales we
conclude that at the moment of recombination the universe was extremely
homogeneous and its temperature could not vary from place to place more than
about few times in thousandth of the percent.

On the other hand, the origin of the large scale structure requires the
presence of small inhomogeneities in the distribution of the matter and
therefore the temperature of CMB should also vary a little bit. These
variations are observed today as the angular fluctuations of the CMB
temperature \cite{WMAP}. The expected fluctuations in a given angular scale
are basically determined by the inhomogeneities on the spatial scales having
today an appropriate angular size if placed at the distance corresponding to
the recombination redshift.

The Hubble scale at recombination epoch plays especially important role,
distinguishing the inhomogeneities which are still frozen from those ones
which already entered the horizon and therefore could be amplified by
gravitational instability. At the scales bigger than the Hubble size, the
perturbations generated during inflation remain unchanged. Therefore,
observing the fluctuations on the angular scales $\theta >1^{\circ }$,
corresponding to super-Hubble scales at recombination, we directly probe the
primordial inflationary spectrum not influenced by the evolution. The
perturbations which entered the horizon before recombination evolve in a
rather complicated way. The transfer functions relating the initial spectrum
to the resulting one strongly depends on the major cosmological parameters,
and the shape of the CMB fluctuations spectrum at $\theta <1^{\circ }$ is
very sensitive to the exact values of these parameters. Therefore by
measuring the fluctuations at small angular scales we can determine these
parameters.

The recent observations of the CMB fluctuations \cite{WMAP} \ give us a hope
that finally we will be able to determine the cosmological parameters with a
very high precision. One of the most important parameters among them is the
spectral index $n_{s}$ characterizing the initial perturbations. According
to inflationary paradigm $n_{s}$ should deviate from $n_{s}=1$ and be in the
range $0.92<n_{s}<0.97$ depending on the particular scenario of the simple%
\footnote{%
Under simple inflation I mean the scenarios with the minimal number of free
parameters.} inflation \cite{MC}, \cite{MB}. It is very important to find
these deviations to confirm or disprove inflationary paradigm. The accuracy
of the current observations is not yet high enough to conclude about the
deviations of the spectral index from the flat one \cite{Efs}. However the
future measurements will be able to reach the needed precision.

The CMB spectrum depends on the various cosmological parameters in a rather
complicated way. It is very important to clarify this dependence to be sure
which features of the spectrum are most sensitive to the particular
combinations of cosmological parameters. The usual approach using the
computer code CMBFAST \cite{SZa} is very helpful, but it does not completely
solve the problem since the parameter space has too many dimensions. There
are various semi-analytical and analytical approaches to this problem \cite%
{others},\cite{W}. However I was not able to find in the literature
elementary analytical expression which would explicitly describe the
dependence of the CMB spectrum on the cosmological parameters and would be
in a reasonably good agreement with numerics. In this paper I derive such
expressions. The main results of this paper are the equations (\ref{Fflt00}%
)-(\ref{Fflt6}).

I start with a pedagogical introduction reminding the derivation of the
Sachs-Wolfe effect in the conformal Newtonian coordinate system \cite{MB}
and first make the calculations assuming the instantaneous recombination. In
this approximation the radiation can be well described in a perfect fluid
approximation before recombination and as an ensemble of free photons
immediately after that. This is well justified by causality only when we
consider the fluctuations corresponding to the superhorizon scales. At small
angular scales the delayed recombination is quite important and leads to an
extra damping of the fluctuations. Therefore as a next step I show how the
formulae obtained in the instantaneous recombination approximation should be
modified to account for this effect. Finally the spectrum for the small
angular scales is derived and the precision of the determination of
cosmological parameters and degeneracy of the spectrum with respect to
certain combinations of these parameters is discussed.

One important simplification I make is that I nearly always consider the
most observationally favored case of a flat universe. The modifications of
the most important features of the CMB spectrum due to the spatial curvature
are rather obvious.

In Appendix A I derive the analytical formulae describing non-instantaneous
recombination which I use in the section on the finite thickness effects. In
Appendix B the transfer functions in short and longwave limits are derived
in the conformal Newtonian gauge.

\section{Sachs-Wolfe effect}

Before recombination the radiation is strongly coupled to the matter and it
can be well described by a perfect fluid approximation. When the hydrogen
becomes neutral, most of the photons do not interact anymore with the matter
and therefore to describe them we need the kinetic equation.

The free propagating photons are described by the distribution function $f$
defined via 
\begin{equation}
dN=f(x^{i},p_{j},\eta )d^{3}xd^{3}p  \label{distfunct}
\end{equation}%
where $dN$ is the number of particles at time $\eta $ in the appropriate
element of the phase volume $d^{3}xd^{3}p\equiv
dx^{1}dx^{2}dx^{3}dp_{1}dp_{2}dp_{3}$, so that $f$ is the particle density
in the one-particle phase space. I assume that the indices $\alpha ,\beta ...
$ run always over $0,...,3$ while $i,k$ take only spatial values $1,2,3.$The
phase volume is invariant under coordinate transformations and hence the
distribution function $f$ \ is a space-time scalar. Since the phase volume
is conserved along the trajectory, the distribution function \textit{in the
absence of the scatterings} should obey the collisionless Boltzmann equation 
\begin{equation}
\frac{Df\left( x^{i}\left( \eta \right) ,p_{i}\left( \eta \right) ,\eta
\right) }{D\eta }\equiv \frac{\partial f}{\partial \eta }+\frac{dx^{i}}{%
d\eta }\frac{\partial f}{\partial x^{i}}+\frac{dp_{i}}{d\eta }\frac{\partial
f}{\partial p_{i}}=0  \label{Boltzmann}
\end{equation}%
where $dx^{i}/d\eta $ and $dp_{i}/d\eta $ are the appropriate derivatives
calculated on the photon's geodesic.

\textit{Temperature and its transformation properties. }The energy
(frequency) of the photon with the 4-momentum $p_{\alpha }$ measured by an
observer having the 4-velocity $u^{\alpha }$ is equal to the scalar product
of these vectors: $\omega =$ $p_{\alpha }u^{\alpha }.$ This can be easily
understood by going to the local inertial coordinate frame of the observer.
If the spectrum of the quanta coming to an observer from a particular
direction on the sky, characterized by the vector $n_{i}=-p_{i}/p,$ where $%
p=\left( \Sigma p_{i}^{2}\right) ^{1/2}$, is the Planckian one then the
temperature, defined via 
\begin{equation}
f=\bar{f}\left( \frac{\omega }{T}\right) \equiv \frac{2}{\exp (\omega
/T(x^{\alpha },n_{i}))-1}  \label{tempdef}
\end{equation}%
generically depends not only on the direction $n_{i}$ but also on the moment
of time $\eta $ and the position of the observer $x^{i}.$ The factor two
here accounts for two possible polarizations of the photons. From now on I
consider the Universe where the fluctuations of the temperature are very
small and therefore one can write 
\begin{equation}
T(x^{\alpha },l_{i})=T_{0}\left( \eta \right) +\delta T(x^{\alpha },n_{i})
\label{tempuniv}
\end{equation}%
where $\delta T$ is much smaller that homogeneous component $T_{0}.$ If the
observer is at rest with respect to a certain coordinate system then taking
into account that $g_{\alpha \beta }u^{\alpha }u^{\beta }=g_{00}\left(
u^{0}\right) ^{2}=1$ we find that the photon frequency measured by this
observer is equal to $\omega =p_{0}/\sqrt{g_{00}}.$ If one goes to the other
coordinate system $\widetilde{x}^{\alpha }=x^{\alpha }+\xi ^{\alpha }$
infinitesimally different from the old one, then the frequency of the same
photon, measured by a different observer, who is at rest with respect to
this new coordinate system, changes. From the transformation properties of
the metric and the 4-momentum one gets that 
\begin{equation}
\omega \Rightarrow \tilde{\omega}=\tilde{p}_{0}/\sqrt{\tilde{g}_{00}}=\omega
(1+\xi ^{i\prime }n_{i})  \label{freqtrans}
\end{equation}%
where I used the eq. $p_{\alpha }p^{\alpha }=0$ and kept only the first
order terms in $\xi $ and metric perturbations; prime denotes the derivative
with respect to time $\eta .$ Taking into account that the distribution
function is a scalar, one easily finds that the small temperature
fluctuations measured by an observer (at rest) in the new coordinate system
are given by 
\begin{equation}
\widetilde{\delta T}=\delta T-T_{0}^{\prime }\xi ^{0}+T_{0}\xi ^{i\prime
}n_{i}  \label{newtemp}
\end{equation}%
Hence, we see that only the monopole and dipole components depend on the
particular coordinate system. The monopole component can always be removed
by a redefinition of the background temperature and can not be measured
locally. The dipole component depends on the motion of the observer with
respect to the \textquotedblright new ether\textquotedblright\ defined by
the background radiation and measuring it we can find how the Earth moves
relative to CMB. Both of these components are not very interesting regarding
the spectrum of the initial fluctuations. The higher multipoles depend
neither on the particular observer or coordinate system we use to calculate
them. Therefore I perform the calculations in conformal Newtonian coordinate
system where these calculations look especially simple.

Let us solve the Boltzmann's equation for the free propagating radiation in
a flat universe with the metric 
\begin{equation}
ds^{2}=a^{2}\left\{ \left( 1+2\Phi \right) d\eta ^{2}-\left( 1-2\Phi \right)
\delta _{ik}dx^{i}dx^{k}\right\}  \label{metric}
\end{equation}%
where $\Phi \ll 1$ is the gravitational potential. Using the geodesic
equations 
\begin{equation}
\frac{dx^{\alpha }}{d\lambda }=p^{\alpha },\mbox{ \ \ \ \ \ \ }\frac{%
dp_{\alpha }}{d\lambda }=\frac{1}{2}\frac{\partial g_{\gamma \delta }}{%
\partial x^{\alpha }}p^{\gamma }p^{\delta },  \label{geodesic1}
\end{equation}%
where $\lambda $ is an affine parameter, the Boltzmann's equation (\ref%
{Boltzmann}) takes the form 
\begin{equation}
\frac{\partial f}{\partial \eta }+n^{i}\left( 1+2\Phi \right) \frac{\partial
f}{\partial x^{i}}+2p\frac{\partial \Phi }{\partial x^{j}}\frac{\partial f}{%
\partial p_{j}}=0.  \label{Boltzmann1}
\end{equation}%
Taking into account that%
\begin{equation}
\omega =p_{0}/\sqrt{g_{00}}=\left( 1+\Phi \right) p/a,  \label{Fom}
\end{equation}%
and using the Planck the ansatz (\ref{tempdef}), (\ref{tempuniv}) one can
easily get that in the lowest order in perturbations the Boltzmann's
equation reduces to 
\begin{equation}
\left( T_{0}a\right) ^{\prime }=0,  \label{homtemp}
\end{equation}%
while the first order terms lead to 
\begin{equation}
\left( \frac{\partial }{\partial \eta }+n^{i}\frac{\partial }{\partial x^{i}}%
\right) \left( \frac{\delta T}{T}+\Phi \right) =2\frac{\partial \Phi }{%
\partial \eta }.  \label{temperatureeq}
\end{equation}%
In the most interesting case when the universe after recombination is
dominated by dust, a nondecaying mode of the gravitational potential is
constant and therefore the right hand side of the equation (\ref%
{temperatureeq}) vanishes. The operator in the left hand side is a total
time derivative and therefore 
\begin{equation}
\left( \frac{\delta T}{T}+\Phi \right) =const,  \label{integral1}
\end{equation}%
along a null geodesics. The influence of the gravitational potential on the
CMB fluctuations is known as Sachs-Wolfe effect. In the case when the
gravitational potential is time dependent the combination $\left( \delta
T/T+\Phi \right) $ is not constant anymore. As it is clear from (\ref%
{temperatureeq}) its change is given by the integral from the partial time
derivative of the potential along geodesics. This effect is usually called
the integrated Sachs-Wolfe effect. If at late stages the universe is
dominated by quintessence , the integrated SW effect can be essential,
changing the resulting amplitudes of the fluctuations by $10\div 20$ percent
in big angular scales $\theta >1^{\circ }$. The accounting of this effect is
rather obvious and therefore to avoid the overcomplication of the final
formulae I consider only the case of the constant potential.

As it follows from the geodesics equations the photons arriving at present
time $\eta _{0}$ to observer located at $x_{0}^{i}$ from the direction $n^{i}
$ propagate along geodesics 
\begin{equation}
x^{i}(\eta )\simeq x_{0}^{i}+n^{i}\left( \eta -\eta _{0}\right) .
\label{integral2}
\end{equation}%
Therefore, from (\ref{integral1}) we get that $\delta T/T$ in the direction $%
n^{i}$ on the sky is equal today to 
\begin{equation}
\frac{\delta T}{T}(\eta _{0},x_{0}^{i},n^{i})=\frac{\delta T}{T}(\eta
_{r},x^{i}(\eta _{r}),n^{i})+\left[ \Phi (\eta _{r},x^{i}(\eta _{r}))-\Phi
\left( \eta _{0},x_{0}^{i}\right) \right]   \label{fluct1}
\end{equation}%
where $\eta _{r}$ is the recombination moment and $x^{i}(\eta _{r})$ is
given by (\ref{integral2}). Since we live in a particular place we are only
interested in $n^{i}$ -dependence of the temperature fluctuations.
Therefore, the last term in (\ref{fluct1}), contributing only to the
monopole component, which is not measurable locally anyway, can be ignored.
As we see the angular dependence of $\left( \delta T/T\right) _{0}$ is
determined by two factors: a) by the \textquotedblright initial
value\textquotedblright\ of $\left( \delta T/T\right) _{r}$ in $\mathbf{n}-$%
direction in a place from where the photons arrive and b) by the value of
the gravitational potential $\Phi $ in this place. The appropriate
temperature fluctuations at the moment of recombination $\left( \delta
T/T\right) _{r}$ can be easily expressed in terms of the gravitational
potential and the fluctuations of the photon energy density $\delta _{\gamma
}\equiv \delta \varepsilon _{\gamma }/\varepsilon _{\gamma }$ at this time.
With this purpose let us use the matching conditions for the hydrodynamical
energy momentum tensor (EMT), which describes the radiation before
decoupling, and the kinetic EMT (see, for instance,\cite{MTW})%
\begin{equation}
T_{\beta }^{\alpha }=\frac{1}{\sqrt{-g}}\int d^{3}pf\frac{p^{\alpha
}p_{\beta }}{p^{0}},  \label{Femtk}
\end{equation}%
characterizing the gas of the free photons after decoupling. Substituting
the expression for the metric into (\ref{Femtk}) and using for the
distribution function ansatz (\ref{tempdef}) we get (up to the linear in
perturbations terms):%
\begin{equation}
T_{0}^{0}\simeq \frac{\left( 1+2\Phi \right) }{a^{4}}\int d^{3}p\bar{f}%
\left( \frac{\omega }{T}\right) p_{0}\simeq T_{0}^{4}\int \left( 1+4\frac{%
\delta T}{T_{0}}\right) \bar{f}\left( y\right) y^{3}dyd^{2}l,  \label{Femt00}
\end{equation}%
where $y\equiv \omega /T$ and we have expressed $p_{0}$ and $p$ through $%
\omega .$ The integral over $y$ from the Planckian function $\bar{f}$ can be
explicitly calculated and give the numerical factor, which, being combined
with $4\pi T_{0}^{4},$ is equal to the energy density of the unperturbed
radiation. Right before recombination the appropriate component of
hydrodynamical EMT for the radiation is equal to $T_{0}^{0}=\varepsilon
_{\gamma }\left( 1+\delta _{\gamma }\right) .$ This component doesn't jump
at the moment when the universe becomes transparent and hence%
\begin{equation}
\delta _{\gamma }=4\int \frac{\delta T}{T}\frac{d^{2}n}{4\pi }.
\label{F00cont}
\end{equation}%
Similar by, one can derive from (\ref{Femtk}) that for the kinetic EMT 
\begin{equation}
T_{0}^{i}\simeq 4\varepsilon _{\gamma }\int \frac{\delta T}{T}n^{i}\frac{%
d^{2}n}{4\pi }.  \label{F0i}
\end{equation}%
On the other hand as it follows from the conservation law for the coupled
photon-baryon plasma (\ref{R10a}) (see appendix B) the appropriate
divergence for the hydrodynamical components of $T_{0}^{i}$ can be expressed
in terms of $\delta _{\gamma }$ and $\Phi ;$ hence 
\begin{equation}
\delta _{\gamma }^{\prime }=-4\int n^{i}\nabla _{i}\left( \frac{\delta T}{T}%
\right) \frac{d^{2}n}{4\pi }.  \label{F0im}
\end{equation}%
where I have assumed that at recombination the cold matter dominates and
therefore neglected the time derivative of the potential: $\Phi ^{\prime
}\left( \eta _{r}\right) =0$. Going to the Fourier space we easily infer that%
\begin{equation}
\left( \frac{\delta T}{T}\right) _{k}\left( \eta _{r}\right) =\frac{1}{4}%
\left( \delta _{k}+\frac{3i}{k^{2}}\left( k_{m}n^{m}\right) \delta
_{k}^{\prime }\right)   \label{Fdt}
\end{equation}%
satisfies both matching conditions (\ref{F00cont}) and (\ref{F0im}). Here
and later on we skip the index $\gamma $ assuming that $\delta $ always
denotes the fluctuations of the radiation energy density. Taking into
account the initial conditions (\ref{Fdt}) and skipping the monopole term in
(\ref{fluct1}), we obtain the following expression for the temperature
fluctuations in the direction $\mathbf{n}\equiv (n^{1},n^{2},n^{3})$ at
location $\mathbf{x}_{0}\equiv (x^{1},x^{2},x^{3})$ 
\begin{equation}
\frac{\delta T}{T}(\eta _{0},\mathbf{x}_{0},\mathbf{n})=\int \left( \left(
\Phi +\frac{\delta }{4}\right) _{\mathbf{k}}-\frac{3\delta _{\mathbf{k}%
}^{\prime }}{4k^{2}}\frac{\partial }{\partial \eta _{0}}\right) _{\eta
_{r}}e^{i\mathbf{k}\cdot \left( \mathbf{x}_{0}+\mathbf{n}\left( \eta
_{r}-\eta _{0}\right) \right) }\frac{d^{3}k}{\left( 2\pi \right) ^{3/2}}
\label{Fflt}
\end{equation}%
where $k\equiv \left\vert \mathbf{k}\right\vert ,$ $\mathbf{k\cdot n}\equiv
k_{m}n^{m}$ and $\mathbf{k\cdot x}_{0}\equiv k_{n}x_{0}^{n}.$ Since $\eta
_{r}/\eta _{0}\lesssim 1/30$ we can neglect here $\eta _{r}$ compared to $%
\eta _{0}.$ It is clear that the first term under the integral represents
the combined result from the initial inhomogeneities in the radiation energy
density and Sachs-Wolfe effect, while the second term is related to the
velocities of the baryon-radiation plasma at recombination and therefore, is
called Doppler contribution to the fluctuations.

\section{Correlation function and multipoles}

In the experiments one usually measures the temperature difference of the
photons received by two antennae separated by a given angle $\theta $ and
this squared difference is averaged over the substantial part of the sky.
The obtained quantity can be expressed in terms of the correlation function 
\begin{equation}
C\left( \theta \right) =\left\langle \frac{\delta T}{T_{0}}\left( \mathbf{n}%
_{1}\right) \frac{\delta T}{T_{0}}\left( \mathbf{n}_{2}\right) \right\rangle
\label{Fcf}
\end{equation}%
where the brackets $\left\langle {}\right\rangle $ denote the averaging over
all $\mathbf{n}_{1}$ and $\mathbf{n}_{2},$ satisfying the condition $\mathbf{%
n}_{1}\bullet \mathbf{n}_{2}=\cos \left( \theta \right) $. Actually, 
\begin{equation}
\left\langle \left( \frac{\delta T}{T_{0}}\left( \theta \right) \right)
^{2}\right\rangle \equiv \left\langle \left( \frac{T\left( \mathbf{n}%
_{1}\right) -T\left( \mathbf{n}_{2}\right) }{T_{0}}\right) ^{2}\right\rangle
=2\left( C\left( 0\right) -C\left( \theta \right) \right)  \label{Fcf1}
\end{equation}%
On the other hand, for a given perturbation spectrum the correlation
function $C\left( \theta \right) $ can be easily expressed in terms of the
expectation values of the Fourier components of the quantities
characterizing these perturbations at the moment of recombination.

The Universe is homogeneous and isotropic in big scales and therefore the
averaging over the sky for a particular observer and a spatial averaging
over the positions $\mathbf{x}_{0}$ should give for small angles (or big
multipoles) nearly the same results. Therefore, the problem of averaging is
finally reduced to the averaging of the products of Fourier components for
the random Gaussian field. Substituting (\ref{Fflt}) into (\ref{Fcf}) and
taking into account that, $\left\langle \Phi _{\mathbf{k}}\Phi _{\mathbf{k}%
^{\prime }}\right\rangle $ $=\left\vert \Phi _{k}\right\vert ^{2}\delta
\left( \mathbf{k+k}^{\prime }\right) ,$ after integrating over the angular
part of $\mathbf{k}$ we obtain: 
\begin{equation}
C=\int \left( \Phi _{k}+\frac{\delta _{k}}{4}-\frac{3\delta _{k}^{\prime }}{%
4k^{2}}\frac{\partial }{\partial \eta _{1}}\right) \left( \Phi _{k}+\frac{%
\delta _{k}}{4}-\frac{3\delta _{k}^{\prime }}{4k^{2}}\frac{\partial }{%
\partial \eta _{2}}\right) ^{\ast }\frac{\sin \left( k\left\vert \mathbf{n}%
_{1}\eta _{1}-\mathbf{n}_{2}\eta _{2}\right\vert \right) }{k\left\vert 
\mathbf{n}_{1}\eta _{1}-\mathbf{n}_{2}\eta _{2}\right\vert }\frac{k^{2}dk}{%
2\pi ^{2}},  \label{Fcf3}
\end{equation}%
where after differentiation with respect to $\eta _{1}$ and $\eta _{2}$ we
have to put $\eta _{1}=\eta _{2}=\eta _{0}.$ Now using the formula (see
(10.1.45) in \cite{AS}):%
\begin{equation}
\frac{\sin \left( k\left\vert \mathbf{n}_{1}\eta _{1}-\mathbf{n}_{2}\eta
_{2}\right\vert \right) }{k\left\vert \mathbf{n}_{1}\eta _{1}-\mathbf{n}%
_{2}\eta _{2}\right\vert }=\sum_{l=0}^{\infty }\left( 2l+1\right)
j_{l}\left( k\eta _{1}\right) j_{l}\left( k\eta _{2}\right) P_{l}\left( \cos
\theta \right)   \label{Fexp}
\end{equation}%
where $P_{l}\left( \cos \theta \right) $ and $j_{l}\left( k\eta \right) $
are, respectively, the Legendre polynomials and spherical Bessel functions
of order $l,$ we can rewrite the expression for the correlation function in
the following form 
\begin{equation}
C\left( \theta \right) =\frac{1}{4\pi }\sum_{l=2}^{\infty }\left(
2l+1\right) C_{l}P_{l}\left( \cos \theta \right)   \label{Fexp1}
\end{equation}%
where the monopole and dipole components ($l=0,1$) were excluded and 
\begin{equation}
C_{l}=\frac{2}{\pi }\int \left\vert \left( \Phi _{k}\left( \eta _{r}\right) +%
\frac{\delta _{k}\left( \eta _{r}\right) }{4}\right) j_{l}\left( k\eta
_{0}\right) -\frac{3\delta _{k}^{\prime }\left( \eta _{r}\right) }{4k}\frac{%
dj_{l}\left( k\eta _{0}\right) }{d\left( k\eta _{0}\right) }\right\vert
^{2}k^{2}dk.  \label{Fmul}
\end{equation}%
The coefficients $C_{l}$ are directly related to the coefficients \ $a_{lm}$
in the expansion of $\delta T/T$ in terms of spherical harmonics, namely $%
C_{l}=\left\langle \left\vert a_{lm}\right\vert ^{2}\right\rangle ,$ and
therefore they characterize the contribution of the multipole component $l$
to the correlation function. If $\theta \ll 1$ the main contribution to $%
C\left( \theta \right) $ give the multipoles with $l\sim 1/\theta .$

The resulting spectrum of CMB-fluctuations depends of the various
cosmological parameters. First of all, these are the amplitude and the index
of the primordial spectrum of inhomogeneities, generated by inflation. The
rather generic prediction of inflation is that in the interesting for us
scales\footnote{%
Most of the calculations will be done here for a flat spectrum ($n=1$). The
consideration can be easily generalized for an arbitrary spectral index $n$
and in the case of small deviations from the flat spectrum the modification
of the final results is obvious (it will be briefly discussed later).}: $%
\left\vert \Phi _{k}^{2}k^{3}\right\vert =Bk^{n-1},$ with $1-n_{s}\sim
0.03\div 0.08$ \cite{MC}, \cite{MB}. The amplitude $B$ is not predicted and
should be normalized to fit the observations. The other parameters on which
the shape of the CMB-spectrum depends are the baryon density, characterized
by $\Omega _{b},$ the contribution of the clustered cold matter to the total
energy density $\Omega _{m}$ ($\Omega _{m}=\Omega _{b}+\Omega _{cdm}$)$,$
the Hubble constant $h_{75}$ (I normalize it on $75$ $km/sec\cdot Mpc$) and
the cosmological constant (quintessence) characterized by $\Omega _{\Lambda
}.$ The present data are best fitted assuming that the universe is flat with 
$\Omega _{tot}=\Omega _{m}+\Omega _{\Lambda }\simeq 1$ and the total energy
density is dominated by the dark cold matter and quintessence with only a
few percent of \ baryons. Below I concentrate mostly on the models, which
deviate from the \textquotedblright concordance model\textquotedblright\ not
too much. Our purpose is to clarify how the variation of the parameters
influences the observed CMB spectrum and to get an idea up to what extent
the CMB determination of the cosmological parameters is robust. I will
consider the different angular scales separately.

\section{ Anisotropies in big angular scales}

The formula (\ref{Fmul}) was derived in the approximation of the
instantaneous recombination. Because of causality this approximation is
rather good when we consider big angular scales, where the CMB fluctuations
are mainly determined by inhomogeneities exceeding the horizon scale at
recombination. Moreover, the perturbations spectrum in these scales is not
much influenced by the evolution. Hence the CMB fluctuations in big angular
scales deliver us the undisturbed information about primordial
inhomogeneities, which are characterized by the amplitude of the primordial
spectrum $B$ and by the spectral index $n_{s}.$ The horizon at recombination
is about the Hubble scale $H_{r}^{-1}=1.5t_{r},$ which in flat universe has
the angular size $0.87^{\circ }$ on today's sky. Therefore, the fluctuations
which we will consider in this section refer to the angles $\theta \gg
1^{\circ }$ or, to the multipoles $l\ll 1/\theta _{H}\sim 200.$

For the superhorizon adiabatic perturbations with $k\eta _{r}\ll 1$ we have
(see Appendix B): 
\begin{equation}
\delta _{k}\left( \eta _{r}\right) \simeq -\frac{8}{3}\Phi _{k},\mbox{ \ \ }%
\delta _{k}^{\prime }\left( \eta _{r}\right) \simeq 0.  \label{Fdelta}
\end{equation}%
As it follows from (\ref{Fflt}) their contribution to the temperature
fluctuations is equal to 
\begin{equation}
\frac{\delta T}{T}(\eta _{0},\mathbf{x}_{0},\mathbf{n})\simeq \frac{1}{3}%
\Phi \left( \eta _{r},\mathbf{x}_{0}-\mathbf{n}\eta _{0}\right) ,
\label{Fonethird}
\end{equation}%
that is the observed fluctuations constitute one third of the gravitational
potential in a place from where the photons arrived. Taking into account
that after equality the potential in supercurvature scales drops compared to
its initial value $\Phi _{k}^{0}$ by factor $9/10$ \cite{MB}, substituting (%
\ref{Fdelta}) into (\ref{Fmul}) and calculating the integral with the help
of the standard formula%
\begin{equation}
\int_{0}^{\infty }s^{m-1}j_{l}^{2}\left( s\right) =2^{m-3}\pi \frac{\Gamma
\left( 2-m\right) \Gamma \left( l+\frac{m}{2}\right) }{\Gamma ^{2}\left( 
\frac{3-m}{2}\right) \Gamma \left( l+2-\frac{m}{2}\right) }  \label{Fb}
\end{equation}%
for the flat initial spectrum ($\left\vert \left( \Phi _{k}^{0}\right)
^{2}k^{3}\right\vert =B$) we obtain well known result:%
\begin{equation}
\left( l\left( l+1\right) C_{l}\right) _{l<30}=\frac{9B}{100\pi }=const.
\label{Fsp}
\end{equation}

Deriving this formula I used in the integrand the flat spectrum everywhere,
assuming that the main contribution for small $l$ comes from the scales
exceeding the horizon, where the primordial spectrum is not modified by
evolution. This is a rather good approximation for $l$ up to $20\div 30$.
Nonetheless, when we are interested in the precise normalization we need to
take into account the corrections coming from the modified spectrum of the
perturbations at big $k.$ This can be well traced only in numerical
calculations.

Unfortunately, the accuracy of the direct information about the statistical
properties of the primordial perturbations spectrum gained from the
measurements in big angular scales is restricted by the \textit{cosmic
variance}. In fact, within cosmic horizon there are only $2l+1$ samples of
the statistical realization for every particular multipole component $l.$
This leads to the minimal inevitable typical \textquotedblright statistical
fluctuations\textquotedblright\ in $C_{l}$%
\begin{equation}
\frac{\Delta C_{l}}{C_{l}}\simeq \left( 2l+1\right) ^{-1/2}.  \label{Fer}
\end{equation}%
Hence, the statistical properties of the spectrum in the scales
corresponding to the multipole $l$ can be determined in observations only up
to an inevitable "error" given by (\ref{Fer}). For the quadrupole ($l=2$)
this \textquotedblright typical error\textquotedblright\ is about $50\%$ and
therefore it can not be used for the normalization of the spectrum. For $%
l\sim 20$ the error constitutes $15\%.$ Therefore, if we want to get a
better accuracy in determining the spectrum of primordial inhomogeneities we
are forced to go to smaller angular scales, where the spectrum is distorted
by evolution. On the one hand it is bad news, since we lose the
\textquotedblright pristine information\textquotedblright . However, on the
other hand, the distortions of the spectrum depend on the other cosmological
parameters involving them \textquotedblright directly in the
game\textquotedblright\ and, therefore, allowing us to determine these
parameters under condition of having precise enough measurements.

On small angular scales we can not ignore anymore the effect of the delayed
recombination and the obtained above formulae should be corrected. Therefore
before I proceed with calculations of the fluctuations in small scales I
will find how the basic formulae should be modified to account for the
effect of delayed recombination.

\section{Delayed recombination and finite thickness effect}

The delayed (non-instantaneous) recombination is important because of two
reasons. First of all, the finite duration of recombination makes the moment
when a specific photon decouples to be not very definite. As a result the
information about the place from where this photon arrives is
\textquotedblright smeared out\textquotedblright . This leads to a
suppression of the CMB-fluctuations in small angular scales, known as finite
thickness effect. The delayed recombination leads also to an extra
dissipation of the inhomogeneities increasing the Silk damping scale and
hence changing the conditions in the places where the photons decouple.
First we consider the finite thickness effect.

Let us consider a particular photon arriving to us from the direction\textbf{%
\ }$\mathbf{n}$. With non-negligible probability this photon could decouple
at any value of the redshift in the interval: $1200>z>900$ and propagate
without further scatterings afterwards. If this happens at the moment $\eta
_{L}$ then the photon arrives to us from the place $\mathbf{x}\left( \eta
_{L}\right) =\mathbf{x}_{0}+\mathbf{n}\left( \eta _{L}-\eta _{0}\right) $
without further scatterings and brings the information about conditions in
this particular place. Since we do not know exactly when and where the
particular photon decouples, a set of the photons arriving from a definite
direction brings us only \textquotedblright smeared\textquotedblright\
information about the conditions within the layer of width $\Delta x\sim
\Delta \eta _{L},$ where $\Delta \eta _{L}$ is the duration of
recombination. It is clear that if the perturbation has a scale smaller than 
$\Delta \eta _{L}$ then as a result of this smearing the information about
the structure of this perturbation will be lost and we expect that the
contribution of these scales to the temperature fluctuations will be
strongly suppressed.

Let us calculate the probability that the photon was scattered last time
within the time interval $\Delta t_{L}$ at the moment of physical time $%
t_{L} $ (corresponding to the conformal time $\eta _{L})$ and then avoided
further scatterings until present time $t_{0}$. With this purpose we divide
the time interval $t_{0}>t>t_{L}$ into $N$ small pieces of duration $\Delta
t,$ so that, $t_{j}=t_{L}+j\Delta t$ and $N>j>1.$ It is obvious that the
required probability is 
\begin{equation}
\Delta P=\frac{\Delta t_{L}}{\tau \left( t_{L}\right) }\left( 1-\frac{\Delta
t}{\tau \left( t_{1}\right) }\right) ...\left( 1-\frac{\Delta t}{\tau \left(
t_{j}\right) }\right) ...\left( 1-\frac{\Delta t}{\tau \left( t_{N}\right) }%
\right) ,  \label{Fp}
\end{equation}%
where $\tau \left( t_{j}\right) =\left( \sigma _{T}n_{t}\left( t_{j}\right)
X\left( t_{j}\right) \right) ^{-1}$ is the mean free time due to the
Thompson scattering at $t_{j}$ and $n_{t}$, $X$ are, respectively, the total
number density of all (bounded and free) electrons and the degree of
ionization. Taking limit $N\rightarrow \infty $ ($\Delta t\rightarrow 0$)
and going back from the physical time $t$ to conformal time $\eta $ we
obtain:%
\begin{equation}
dP\left( \eta _{L}\right) =\mu ^{\prime }\left( \eta _{L}\right) \exp \left(
-\mu \left( \eta _{L}\right) \right) d\eta _{L},  \label{Fp1}
\end{equation}%
where prime, as usually, denotes the derivative with respect to conformal
time and the optical depth%
\begin{equation}
\mu \left( \eta _{L}\right) \equiv \int_{t_{L}}^{t_{0}}\frac{dt}{\tau \left(
t\right) }=\int_{\eta _{L}}^{\eta _{0}}\sigma _{T}n_{t}X_{e}a\left( \eta
\right) d\eta .  \label{Fp3}
\end{equation}%
was introduced. Now, taking into account that in the formula (\ref{Fflt})
the recombination moment $\eta _{r}$ should be replaced by $\eta _{L}$
weighted with the probability (\ref{Fp1},) we conclude that this formula
should be modified as:%
\begin{equation}
\frac{\delta T}{T}=\int \left\{ \Phi +\frac{\delta }{4}-\frac{3\delta
^{\prime }}{4k^{2}}\frac{\partial }{\partial \eta _{0}}\right\} _{\eta
_{L}}e^{i\mathbf{k}\cdot \left( \mathbf{x}_{0}+\mathbf{n}\left( \eta
_{L}-\eta _{0}\right) \right) }\mu ^{\prime }\exp \left( -\mu \right) d\eta
_{L}\frac{d^{3}k}{\left( 2\pi \right) ^{3/2}}  \label{Fp4}
\end{equation}%
I would like to stress that in distinction from (\ref{Fflt}) here one can
not neglect $\eta _{L}$ compared to $\eta _{0}$ anymore since when we
integrate over $\eta _{L}$ the appropriate argument of the exponent changes
very much for $k>\eta _{L}^{-1}$.

It is easy to see that the visibility function $\mu ^{\prime }\exp \left(
-\mu \right) $ vanishes at very small $\eta _{L}$ (because $\mu \gg 1$) and
at big $\eta _{L}$ ($\mu ^{\prime }\rightarrow 0$) and reaches the maximum
at $\eta _{r}$ determined by the condition%
\begin{equation}
\mu ^{\prime \prime }=\mu ^{\prime 2}  \label{Fp5}
\end{equation}%
Since in the case of non-instantaneous recombination the moment when the
photons decouple from the matter become smeared over rather substantial time
interval we reserve from now on the notation $\eta _{r}$ for the time when
the visibility function takes its maximum value. This maximum is located
within thin layer $1200>z>900$. During this short time interval the scale
factor and the total number density $n_{t}$ do not change very substantially
and therefore we neglect their time dependence, estimating the appropriate
values at $\eta =\eta _{r}$. On the contrary, the ionization degree $X$
changes by few orders of magnitude. Taking this into account we can rewrite
the condition (\ref{Fp5}) as:%
\begin{equation}
X_{r}^{\prime }\simeq -\left( \sigma _{T}n_{t}a\right) _{r}X_{r}^{2}
\label{Fp6}
\end{equation}%
where index $r$ means that the appropriate quantities are estimated at $\eta
_{r}$. At $1200>z>900$ the ionization degree $X$ is well described by the
formula (\ref{Hsol2}) in Appendix A. The change of $X$ is mainly due to the
exponential factor there; hence%
\begin{equation}
X^{\prime }\simeq -\frac{1.44\times 10^{4}}{z}\mathcal{H}X  \label{Fp7}
\end{equation}%
where $\mathcal{H}\equiv \left( a^{\prime }/a\right) .$ Substituting this
relation in (\ref{Fp6}) we get%
\begin{equation}
X_{r}\simeq \mathcal{H}_{r}\kappa \left( \sigma _{T}n_{t}a\right) _{r}^{-1}
\label{Fp8}
\end{equation}%
where $\kappa \equiv 14400/z_{r}.$ Together with (\ref{Hsol2}) this equation
determines when the visibility function takes its maximum value. It is easy
to see that this happens in the "middle" of the recombination layer at $%
z_{r}\simeq 1050$ irrespective of the values of the cosmological parameters.
At this time the ionization degree $X_{r}$ is still $\kappa \simeq 13.7$
times bigger than the ionization degree at the moment of decoupling
determined by condition $t\sim \tau _{\gamma }$ (see (\ref{Hxdec})). Near
its maximum the visibility function can be well approximated by the Gaussian
one:

\begin{equation}
\mu ^{\prime }\exp \left( -\mu \right) \propto \exp \left( -\frac{1}{2}%
\left( \mu -\ln \mu ^{\prime }\right) _{r}^{\prime }(\eta _{L}-\eta
_{r})^{2}\right)  \label{Fapp1}
\end{equation}%
Calculating the derivatives with the help of (\ref{Fp7}), (\ref{Fp8}) we
obtain%
\begin{equation}
\mu ^{\prime }\exp \left( -\mu \right) \simeq \frac{\left( \kappa \mathcal{H}%
\eta \right) _{r}}{\sqrt{2\pi }\eta _{r}}\exp \left( -\frac{1}{2}\left(
\kappa \mathcal{H}\eta \right) _{r}^{2}\left( \frac{\eta _{L}}{\eta _{r}}%
-1\right) ^{2}\right)  \label{Fp10}
\end{equation}%
where the pre-exponential factor was taken to satisfy the normalization
condition $\int \mu ^{\prime }\exp \left( -\mu \right) d\eta _{L}=1.$

We can use this formula to perform the explicit integration over $\eta _{L}$
in (\ref{Fp4}). The gravitational potential and the slowly varying
contribution to $\delta _{\gamma }$ practically do not change during
recombination. Therefore, they can be approximated by their values at $\eta
_{r}.$ The only term inside the curly brackets in (\ref{Fp4}) which could
incur a very substantial change is the Silk damping scale. Keeping in mind
that the main contribution to the integral comes from the region near $\eta
_{r}$ we estimate this scale also at $\eta _{r}$. Of course this is a rather
rough estimate which nevertheless reproduces the results of the numerics
with rather good accuracy . Thus, ignoring $\eta _{L}-$dependence of the
expression in curly brackets in (\ref{Fp4}) and taking its value at $\eta
_{r}$, after substitution (\ref{Fp10}) in (\ref{Fp4}) and integration over $%
\eta _{L}$ we obtain%
\begin{equation}
\frac{\delta T}{T}=\int \left\{ \Phi +\frac{\delta }{4}-\frac{3\delta
^{\prime }}{4k^{2}}\frac{\partial }{\partial \eta _{0}}\right\} _{\eta
_{r}}\exp \left( -\left( \sigma k\eta _{r}\right) ^{2}\right) e^{i\mathbf{k}%
\cdot \left( \mathbf{x}_{0}+\mathbf{n}\left( \eta _{r}-\eta _{0}\right)
\right) }\frac{d^{3}k}{\left( 2\pi \right) ^{3/2}}  \label{Fp11}
\end{equation}%
where%
\begin{equation}
\sigma \equiv \frac{1}{\sqrt{6}\left( \kappa \mathcal{H}\eta \right) _{r}}
\label{Fp12a}
\end{equation}%
In deriving (\ref{Fp11}) I replaced $\left( \mathbf{k}\cdot \mathbf{n}%
\right) ^{2}$ by $k^{2}/3$, keeping in mind the isotropy of the
perturbations. Note that now we can neglect in the exponent $\eta _{r}$
compared to $\eta _{0}$ . To find how $\sigma $ depends on the cosmological
parameters we have to calculate $\left( \mathcal{H}\eta \right) _{r}.$ At
recombination and before the cosmological term can be ignored and the
behavior of the scale factor is well described by solution for the
matter-radiation universe (see, for instance, \cite{MB}), 
\begin{equation}
a\left( \eta \right) =a_{m}\left( \left( \frac{\eta }{\eta _{\ast }}\right)
^{2}+2\left( \frac{\eta }{\eta _{\ast }}\right) \right) ,  \label{dustradsol}
\end{equation}%
Note, that $\eta _{\ast }$ is a bit different from the equality time $\eta
_{eq}$ when the energy densities of radiation and matter are exactly equal.
The relation $\eta _{eq}$ $=\left( \sqrt{2}-1\right) \eta _{\ast }$ follows
from the equation $a\left( \eta _{eq}\right) =a_{m}$. Hence 
\begin{equation}
\left( \mathcal{H}\eta \right) _{r}=2\frac{1+\left( \eta _{r}/\eta _{\ast
}\right) }{2+\left( \eta _{r}/\eta _{\ast }\right) }  \label{Fp13}
\end{equation}%
where $\left( \eta _{r}/\eta _{\ast }\right) $ can be expressed through the
ratio of redshifts at equality and recombination if we use obvious relation%
\begin{equation}
\left( \frac{\eta _{r}}{\eta _{\ast }}\right) ^{2}+2\left( \frac{\eta _{r}}{%
\eta _{\ast }}\right) \simeq \frac{z_{eq}}{z_{r}}  \label{Fp14}
\end{equation}%
Substituting (\ref{Fp13}) in (\ref{Fp12a}) and taking into account that $%
\kappa \simeq 13.7$ we obtain: 
\begin{equation}
\sigma \simeq 1.49\times 10^{-2}\left( 1+\left( 1+\frac{z_{eq}}{z_{r}}%
\right) ^{-1/2}\right)   \label{Fp17}
\end{equation}%
The exact value of $z_{eq}$ depends on the cold matter contribution to the
total energy density and from the number of the ultrarelativistic species.
Assuming three types of neutrino $z_{r}/z_{eq}$ can be estimated as%
\begin{equation}
\frac{z_{eq}}{z_{r}}\simeq 12.8\left( \Omega _{m}h_{75}^{2}\right) 
\label{Fp15}
\end{equation}%
The value of $\sigma $ depends on the amount of the cold matter not very
sensitively; if $\Omega _{m}h_{75}^{2}\simeq 0.3$ then $\sigma \simeq
2.2\times 10^{-2},$ while for $\Omega _{m}h_{75}^{2}\simeq 1,$ $\sigma
\simeq 1.9\times 10^{-2}.$

Now let us find how non-instantaneous recombination influences the Silk
dissipation scale. As we mentioned above, at $\eta =\eta _{r}$ the
ionization degree is $\kappa \sim 13.7$ times bigger than at the decoupling
and the mean free path of the photon is correspondingly smaller than the
horizon scale. Therefore we can try to use the result (\ref{Rdiss}) of
Appendix B, obtained in imperfect fluid approximation, to estimate the
corrections to the formula (\ref{Rd1}) due to noninstantaneous
recombination. Using the approximate formula (\ref{Hsol2}) which is valid
when the ionization drops below unity we obtain

\begin{equation}
\left( k_{D}\eta \right) _{r}^{-2}\simeq 0.36\left( \Omega
_{m}h_{75}^{2}\right) ^{1/2}\left( \Omega _{b}h_{75}^{2}\right)
^{-1}z_{r}^{-3/2}+\frac{12}{5}c_{S}^{2}\sigma ^{2}  \label{Fp12}
\end{equation}%
The first term here is the same as dissipation scale (\ref{Rd1}) derived for
the case of instantaneous recombination. It accounts the dissipation until
the moment when recombination begins. The second term is due to an extra
dissipation which happens in the process of recombination. Note that the
second term in (\ref{Fp12}) corresponds to the scale which at $\eta _{r}$
smaller that the mean free path $\tau _{\gamma }$ and barely can be trusted
literally. However, within the time interval $\Delta \eta \sim \eta
_{r}\sigma $ when the visibility function is different from zero the free
propagating photons have enough time only to go at the (comoving) distance $%
\lambda \sim \eta _{r}\sigma $ which roughly corresponds to the second term
in (\ref{Fp12}). Hence, although the imperfect fluid interpretation of the
second term becomes questionable, it can be nevertheless used to make an
estimate of the damping scale. For the realistic values of the dark matter
and baryon densities, $\ \Omega _{m}h_{75}^{2}\simeq 0.3$ and $\Omega
_{b}h_{75}^{2}\simeq 0.04,$ this term is nearly twice bigger than the first
term; hence the extra Silk dissipation due to delayed recombination is
rather important. At very low baryon density the first term in (\ref{Fp12})
dominates and most of the dissipation happens before ionization
significantly drops.

Thus, we found that the delayed recombination can be taken into account in a
simple way. First, there occurs the extra dissipation of the perturbations
and the dissipation scale can increase in few times compared to the case of
instantaneous recombination. Second, it leads to an uncertainty when the
photons decouple from the matter and as a result to an extra suppression of
the CMB-fluctuations in small angular scales. Although both effects are
interconnected they have different nature and should not be confused.

The formulae derived in the approximation of instantaneous recombination are
modified in an obvious way. Namely, the formula (\ref{Fflt}) should be
replaced by (\ref{Fp11}). Repeating the steps which lead to the key formula (%
\ref{Fmul}) we conclude that the expression under the integral there should
be just multiplied by a general factor $\exp \left( -2\left( \sigma k\eta
_{r}\right) ^{2}\right) .$

\section{Small angular scales}

At big $l,$ corresponding to small angular scales, the main contribution to $%
C_{l}$ give the perturbations which being placed at recombination have an
angular size $\theta \sim 1/l$ on today's sky . The multipole moment $l\sim
200$ corresponds to the sound horizon scale at recombination. Hence the
perturbations responsible for the fluctuations with $l>100\div 200$ should
have the wavenumbers $k>\eta _{r}^{-1}$, that is, they enter horizon before
recombination. These perturbations evolve in a rather complicated way and
the primordial spectrum is strongly modified at $k>\eta _{r}^{-1}.$ In
realistic models, the transfer functions relating the initial spectrum of
gravitational potential $\Phi _{k}^{0}$ with the resulting spectra for $\Phi 
$ and $\delta _{\gamma }$ at $\eta _{r}$ can be analytically derived only in
two limiting cases: a) for the perturbations which entered horizon \textit{%
well} before equality and b) \textit{much later} after equality (when the
gravitational field of radiation can be ignored). In the limit of very big $%
k $ the result is given by the formulae (\ref{Rd6a}), (\ref{Rd7}), while for
very small $k$ by (\ref{R1com}) (see Appendix B). Unfortunately, for the
realistic values of the cosmological parameters none of these results can be
directly used to calculate the CMB fluctuations in the most interesting
region of first few acoustic peaks. Actually, the derived shortwave
asymptotic is applicable only for those perturbations which have chance for
at least one oscillation before equality ($k\eta _{eq}>2\sqrt{3}\pi \sim 10$)%
$.$ At the same time the longwave asymptotic (\ref{R1com}) can be literally
applied only to the perturbations which entered the horizon when the
radiation was already negligible compared to the matter. If $\Omega
_{m}h_{75}^{2}\simeq 0.3$ then as it follows from (\ref{Fp15}) $%
z_{r}/z_{eq}\simeq 4$, and the radiation still constitute about $20\%$ of
the energy density at the recombination time. Hence, the formula (\ref{R1com}%
) is not trustable for those perturbations which enter the horizon in
between equality and recombination and responsible for the fluctuations in
the region of first few acoustic peaks.

\subsection{Transfer functions}

To describe the perturbations in these intermediate region we have to modify
the derived formulae. Taking into account the time behavior of the
asymptotic WKB-solutions of Appendix B we conclude that at the moment of
recombination: 
\begin{equation}
\Phi _{k}+\frac{\delta _{k}}{4}\simeq \left[ T_{p}\left( 1-\frac{1}{%
3c_{s}^{2}}\right) +T_{o}\sqrt{c_{s}}\cos \left( k\int_{0}^{\eta
_{r}}c_{s}d\eta \right) e^{-\left( k/k_{D}\right) ^{2}}\right] \Phi _{k}^{0}
\label{Fg}
\end{equation}%
and respectively 
\begin{equation}
\delta _{k}^{\prime }\simeq -4T_{o}kc_{s}^{3/2}\sin \left( k\int_{0}^{\eta
_{r}}c_{s}d\eta \right) e^{-\left( k/k_{D}\right) ^{2}}\Phi _{k}^{0}
\label{Fg1}
\end{equation}%
where the transfer functions $T_{p}$ and $T_{o}$ should depend on the
wavenumber $k,$ equality time $\eta _{eq}$ and baryon density $\Omega _{b}$.
To simplify the consideration we will restrict ourselves by the case when
the baryon density is small compared to the total density of the cold
matter, that is, $\Omega _{b}\ll \Omega _{m}.$ This will allow us to neglect
the baryon contribution to the gravitational potential compared to the
contribution of the cold \textit{dark} matter, which interacts with the
radiation only gravitationally. However even in this case the baryons
influence the speed of sound and we have to take this into account. This is
the situation for the concordance model and one can use analytical results,
which I will derived below, only to study the dependence of the fluctuations
on the values of the major cosmological parameters within some
\textquotedblright window\textquotedblright\ around this model. If the
contribution of the baryons to the gravitational potential is negligible the
transfer functions $T_{p}$ and $T_{o}$ depend only on $k$ and $\eta _{eq},$
which on dimensional grounds can enter $T_{p}$ and $T_{o}$ only in
combination $k\eta _{eq}.$ Their asymptotic behavior can be easily inferred
from (\ref{R1com}), (\ref{Rd7}). For the longwave perturbations with $k\eta
_{eq}\ll 1,$ 
\begin{equation}
T_{p}\rightarrow \frac{9}{10};\mbox{ \ \ }T_{o}\rightarrow \frac{9}{10}\cdot
3^{-3/4}\simeq 0.4,  \label{Fg2}
\end{equation}%
while in the shortwave limit for $k\eta _{eq}\gg 1$ 
\begin{equation}
T_{p}\rightarrow \frac{\ln \left( 0.15k\eta _{eq}\right) }{\left( 0.27k\eta
_{eq}\right) ^{2}}\rightarrow 0;\mbox{ \ \ }T_{o}\rightarrow \frac{3^{5/4}}{2%
}\simeq 1.97,  \label{Fg3}
\end{equation}%
where the factor $10/9$ accounts for the change of the gravitational
potential for superhorizon perturbations after matter-radiation equality.
Unfortunately, in the most interesting intermediate range of scales $1<k\eta
_{eq}<10$ which is responsible for the fluctuations in the region of first
few acoustic peaks, the transfer functions can be calculated only
numerically. In the interval: $1<k\eta _{eq}<10,$ one can approximate $T_{p}$
with good accuracy by\cite{BB} 
\begin{equation}
T_{p}\simeq 0.25\ln \left( \frac{14}{k\eta _{eq}}\right) ,  \label{Ftf1}
\end{equation}%
$\,$and, respectively\footnote{%
I would like to thank A. Makarov for numerical calculations of $T_{o}$
function in the limit of vanishing baryon density.}, 
\begin{equation}
T_{o}\simeq 0.36\ln \left( 5.6k\eta _{eq}\right) .  \label{Ftf2}
\end{equation}%
The transfer functions are monotonic; as $k\eta _{eq}$ increases the
function $T_{p}$ decreases and approaches zero, while $T_{o}$ increases and
reaches its asymptotic value $T_{o}\simeq 1.97$. For perturbations which
enter horizon well before equality, the function $T_{o}$ is about five times
bigger than for the perturbations which cross the horizon late after
equality. The physical origin of this difference is rather transparent.
Before equality the gravitational field of the radiation can not be
neglected. Therefore when perturbation enters horizon the gravity field of
the radiation extra boosts the generated sound wave and its amplitude will
be five times bigger than the amplitude in the case when the radiation can
be neglected.

\subsection{Calculating the spectrum}

To calculate $C_{l}$ we should substitute (\ref{Fg}), (\ref{Fg1}) into
formula (\ref{Fmul}), which should be appropriately corrected for the finite
thickness effect. However, the obtained integrals are not very transparent
and before we proceed with their calculation, it makes sense to simplify
these integrals using the advantage of considering $l\gg 1$. With this
purpose we first get rid of the derivatives of the spherical Bessel function
in (\ref{Fmul}). Using the Bessel function equation one can easily verify
that 
\begin{equation}
j_{l}^{\prime 2}\left( z\right) =\left[ 1-\frac{l\left( l+1\right) }{z^{2}}%
\right] j_{l}^{2}\left( z\right) +\frac{\left( zj_{l}^{2}\left( z\right)
\right) ^{\prime \prime }}{2z}  \label{Fbfr}
\end{equation}%
where prime denotes the derivative with respect to the Bessel function
argument. Substituting this into (\ref{Fmul}) and integrating by parts we
get 
\begin{equation}
C_{l}=\frac{2}{\pi }\int \left( \left\vert \Phi +\frac{\delta }{4}%
\right\vert ^{2}k^{2}+\frac{9\left\vert \delta ^{\prime }\right\vert ^{2}}{16%
}\left( 1-\frac{l\left( l+1\right) }{\left( k\eta _{0}\right) ^{2}}\right)
\right) \left( 1+O\right) e^{-2\left( \sigma k\eta _{r}\right)
^{2}}j_{l}^{2}\left( k\eta _{0}\right) dk  \label{Fclsim1}
\end{equation}%
where by $O$ I denoted the corrections of order $\eta _{r}/\eta _{0}$ and $%
\left( k\eta _{0}\right) ^{-1},$ which were estimated taking into account
the general structure of the expressions in (\ref{Fg}), (\ref{Fg1}). The
corrections $\eta _{r}/\eta _{0}$ can be neglected compared to unity since $%
\eta _{r}/\eta _{0}\lesssim z_{r}^{-1/2}\sim 1/30.$ At big $l$ only those $k$
give a substantial contribution to the integral for which $k\eta _{0}\geq l.$
Actually, as $l\rightarrow \infty $ we can use the following approximation
for the Bessel functions%
\begin{equation}
j_{l}\left( z\right) \rightarrow \left\{ 
\begin{array}{cc}
0, & z<\nu , \\ 
z^{-1/2}\left( z^{2}-\nu ^{2}\right) ^{-1/4}\cos \left( \sqrt{z^{2}-\nu ^{2}}%
-\nu \arccos \left( \nu /z\right) -\pi /4\right) , & z>\nu ,%
\end{array}%
\right.  \label{Fbfapp}
\end{equation}%
where $\nu /z\neq 1$ is held fixed and $\nu \equiv l+1/2;$ hence the
correction $1/k\eta _{0}\sim 1/l\ll 1$ can also be skipped.

Now I will use (\ref{Fbfapp}) in the integrand of (\ref{Fclsim1}). Keeping
in mind that the argument of $j_{l}^{2}\left( k\eta _{0}\right) $ changes
with $k$ much faster that the argument of the oscillating part of the WKB
solutions for $\delta _{k}$ let us replace the cosine squared, coming from (%
\ref{Fbfapp}), by its average value $1/2.$ The result reads%
\begin{equation}
C_{l}\simeq \frac{1}{16\pi }\int_{l\eta _{0}^{-1}}^{\infty }\left( \frac{%
\left\vert 4\Phi +\delta \right\vert ^{2}k^{2}}{\left( k\eta _{0}\right) 
\sqrt{\left( k\eta _{0}\right) ^{2}-l^{2}}}+\frac{9\sqrt{\left( k\eta
_{0}\right) ^{2}-l^{2}}}{\left( k\eta _{0}\right) ^{3}}\delta _{k}^{\prime
2}\right) e^{-2\left( \sigma k\eta _{r}\right) ^{2}}dk,  \label{Fcl1}
\end{equation}%
where using the advantage of considering only big multipoles I replaced $l+1$
with $l.$ This result was first derived in \cite{W}.

Let us consider the flat initial spectrum: $\left\vert \left( \Phi
_{k}^{0}\right) ^{2}k^{3}\right\vert =B.$ Substituting (\ref{Fg}), (\ref{Fg1}%
) into (\ref{Fcl1}) and changing the integration variable to $x\equiv k\eta
_{0}/l$\ after elementary calculations we arrive to the following result:%
\begin{equation}
l^{2}C_{l}\simeq \frac{B}{\pi }\left( O+N\right)  \label{Ffcl}
\end{equation}%
where keeping in mind $l-$dependence of $l^{2}C_{l}$ I have written it as a
sum of different terms. Namely,%
\begin{equation}
O\equiv O_{1}+O_{2},  \label{Fo}
\end{equation}%
is the oscillating contribution to the spectrum given by two terms with
twice different periods: 
\begin{equation}
O_{1}=2\sqrt{c_{s}}\left( 1-\frac{1}{3c_{s}^{2}}\right) \int_{1}^{\infty }%
\frac{T_{p}T_{o}e^{\left( -\frac{1}{2}\left( l_{f}^{-2}+l_{S}^{-2}\right)
^{2}l^{2}x^{2}\right) }\cos \left( l\varrho x\right) }{x^{2}\sqrt{x^{2}-1}}%
dx,  \label{Fclo1}
\end{equation}%
and%
\begin{equation}
O_{2}=\frac{c_{s}}{2}\int_{1}^{\infty }T_{o}^{2}\frac{\left(
1-9c_{s}^{2}\right) x^{2}+9c_{s}^{2}}{x^{4}\sqrt{x^{2}-1}}e^{-\left(
l/l_{S}\right) ^{2}x^{2}}\cos \left( 2l\varrho x\right) dx,  \label{Fclo2}
\end{equation}%
These terms modulate the spectrum, leading to the peaks and valleys. I have
introduced here the ratio 
\begin{equation}
\varrho \equiv \frac{1}{\eta _{0}}\int_{0}^{\eta _{r}}c_{s}\left( \eta
\right) d\eta ,  \label{Fshad}
\end{equation}%
which determines the period of oscillations and location of the peaks. The
scales $l_{f}$ and $l_{S}$ characterizing the damping of the fluctuations
because of the Silk dissipation and finite thickness effect are equal to:%
\begin{equation}
l_{f}^{-2}\equiv 2\sigma ^{2}\left( \frac{\eta _{r}}{\eta _{0}}\right) ^{2};%
\mbox{ \ \ }l_{S}^{-2}\equiv 2\left( \sigma ^{2}+\left( k_{D}\eta
_{r}\right) ^{-2}\right) \left( \frac{\eta _{r}}{\eta _{0}}\right) ^{2},
\label{Fd}
\end{equation}%
where $\sigma $ is given in (\ref{Fp17}). The analytical estimate for the
Silk scale $k_{D}\eta _{r}$ is not very accurate, however one still can use
the estimate (\ref{Fp12}) for $k_{D}\eta _{r}$.

In turn, the nonoscillating contribution $I_{c}$ can be written as a sum of
three integrals%
\begin{equation}
N=N_{1}+N_{2}+N_{3},  \label{Fc}
\end{equation}%
where 
\begin{equation}
N_{1}=\left( 1-\frac{1}{3c_{s}^{2}}\right) ^{2}\int_{1}^{\infty }\frac{%
T_{p}^{2}e^{-\left( l/l_{f}\right) ^{2}x^{2}}}{x^{2}\sqrt{x^{2}-1}}dx
\label{FN1}
\end{equation}%
is proportional to the baryon density and vanishes in the absence of baryons
when $c_{s}^{2}=1/3$. The other two integrals are:%
\begin{equation}
N_{2}=\frac{c_{s}}{2}\int_{1}^{\infty }\frac{T_{o}^{2}e^{-\left(
l/l_{S}\right) ^{2}x^{2}}}{x^{2}\sqrt{x^{2}-1}}dx,  \label{FN2}
\end{equation}%
and%
\begin{equation}
N_{3}=\frac{9c_{s}^{3}}{2}\int_{1}^{\infty }T_{o}^{2}\frac{\sqrt{x^{2}-1}}{%
x^{4}}e^{-\left( l/l_{S}\right) ^{2}x^{2}}dx.  \label{FN3}
\end{equation}

Before we proceed further with the calculation of the integrals let us
express the parameters entering (\ref{Ffcl}), namely, $c_{s},l_{f},l_{S},%
\varrho $ and transfer functions $T_{o},T_{p}$ through the basic
cosmological parameters $\Omega _{b},\Omega _{m},h_{75}$ and $\Omega
_{\Lambda }=1-\Omega _{m}.$

\subsection{Parameters}

\textit{The speed of sound} $c_{s}$ at recombination depends only on the
baryon density, which determines how it deviates from the speed of sound in
purely ultrarelativistic medium. To characterize these deviations it is
convenient instead of the baryon density to introduce the parameter $\xi $
defined as 
\begin{equation}
\xi \equiv \frac{1}{3c_{s}^{2}}-1=\frac{3}{4}\left( \frac{\varepsilon _{b}}{%
\varepsilon _{\gamma }}\right) _{r}\simeq 17\left( \Omega
_{b}h_{75}^{2}\right) ,  \label{Fbr1}
\end{equation}%
Then $c_{s}^{2}$ can be expressed through $\xi $ as 
\[
c_{s}^{2}=\frac{1}{3\left( 1+\xi \right) }
\]%
For the realistic value of the baryon density $\Omega _{b}h_{75}^{2}\simeq
0.035$ one gets $\xi \simeq 0.6.$

\textit{The damping scales }$l_{f},l_{S}$ are given by (\ref{Fd}). It is
clear that to express them through the cosmological parameters we first have
to calculate the ratio $\eta _{r}/\eta _{0},$ which also depends on the
cosmological term. To calculate this ratio let us consider an auxiliary
moment of time $\eta _{0}>\eta _{x}>\eta _{r},$ when the radiation is
already negligible and the cosmological term is still not relevant for
dynamics. Then to determine $\eta _{x}/\eta _{0}$ we can use the exact
solution describing a flat universe filled by the matter and cosmological
constant:%
\begin{equation}
a\left( t\right) =a_{0}\left( \sinh \frac{3}{2}H_{0}t\right) ^{2/3}
\label{ddes}
\end{equation}%
As a result we obtain: 
\begin{equation}
\eta _{x}/\eta _{0}\simeq I_{\Lambda }z_{x}^{-1/2}=3\left( \frac{\Omega
_{\Lambda }}{\Omega _{m}}\right) ^{1/6}\left( \int_{0}^{y}\frac{dx}{\left(
\sinh x\right) ^{2/3}}\right) ^{-1}z_{x}^{-1/2},  \label{FI}
\end{equation}%
with the upper limit of integration $y\equiv \sinh ^{-1}\left( \Omega
_{\Lambda }/\Omega _{m}\right) ^{1/2}.$ Taking into account that $\Omega
_{\Lambda }=1-\Omega _{m}$ one can use the following numerical fit for $%
I_{\Lambda }$ in (\ref{FI}): 
\begin{equation}
I_{\Lambda }\simeq \Omega _{m}^{-0.09}  \label{FI1}
\end{equation}%
which approximates the exact result with the accuracy better that $1\%$
everywhere within the interval $0.1<\Omega _{m}<1.$

\bigskip The ratio $\eta _{x}/\eta _{r}$ can be calculated with the help of (%
\ref{dustradsol}) and is equal to 
\begin{equation}
\frac{\eta _{r}}{\eta _{x}}\simeq \left( \frac{z_{x}}{z_{r}}\right)
^{1/2}\left( 1+2\frac{\eta _{\ast }}{\eta _{r}}\right) ^{-1/2}=\left( \frac{%
z_{x}}{z_{eq}}\right) ^{1/2}\left( \left( 1+\frac{z_{eq}}{z_{r}}\right)
^{1/2}-1\right) ,  \label{FI3}
\end{equation}%
where we used the equation (\ref{Fp14}) to express $\eta _{\ast }/\eta _{r}$
in terms of $z_{eq}/z_{r}.$ Combining this formula with (\ref{FI}) we obtain 
\begin{equation}
\frac{\eta _{r}}{\eta _{0}}=\frac{1}{\sqrt{z_{r}}}\left( \left( 1+\frac{z_{r}%
}{z_{eq}}\right) ^{1/2}-\left( \frac{z_{r}}{z_{eq}}\right) ^{1/2}\right)
I_{\Lambda }  \label{FI4}
\end{equation}%
Substituting this together with the expression (\ref{Fp17}) for $\sigma $
into (\ref{Fd}) one gets%
\begin{equation}
l_{f}\simeq 1530\left( 1+\frac{z_{r}}{z_{eq}}\right) ^{1/2}I_{\Lambda }^{-1}
\label{FI5}
\end{equation}%
where the ratio of the redshifts at recombination and equality for three
neutrino types (see (\ref{Fp15})) is equal to%
\begin{equation}
\frac{z_{r}}{z_{eq}}\simeq 7.8\times 10^{-2}\left( \Omega
_{m}h_{75}^{2}\right) ^{-1}  \label{FI6}
\end{equation}%
The scale $l_{f}$ characterizes the damping of CMB-fluctuations because of
finite thickness effect. It depends on both cosmological term and $\Omega
_{m}h_{75}^{2}$ not very sensitively; for instance, if $\Omega
_{m}h_{75}^{2}\simeq 0.3$ and $\Omega _{\Lambda }h_{75}^{2}\simeq 0.7$ we
have $l_{f}\simeq 1580,$ while for $\Omega _{m}h_{75}^{2}\simeq 1$ and $%
\Omega _{\Lambda }h_{75}^{2}\simeq 0$ one gets $l_{f}\simeq 1600.$

The scale $l_{S}$ describing the combined effect from the finite thickness
and Silk damping can be calculated similar by. Using the estimate (\ref{Fp12}%
) for Silk dissipation scale one can easily find that 
\begin{equation}
l_{S}\simeq 0.7l_{f}\left( \frac{1+0.56\xi }{1+\xi }+\frac{0.8}{\xi \left(
1+\xi \right) }\frac{\left( \Omega _{m}h_{75}^{2}\right) ^{1/2}}{\left(
1+\left( 1+z_{eq}/z_{r}\right) ^{-1/2}\right) ^{2}}\right) ^{-1/2}
\label{FI7}
\end{equation}

\bigskip

This formula is not as reliable as the estimate for $l_{f}$ since first we
neglected the contribution of the heat conductivity to Silk dissipation
scale and second we calculated it using imperfect fluid approximation which
surely breaks down when the visibility function reaches its maximum.
Nevertheless it is still trustable within $10\%$ accuracy and an exact
result is a bit smaller than given by (\ref{FI7}). In distinction from $%
l_{f} $ the damping scale $l_{S}$ depends not only on the matter density and
cosmological term but also on the baryon density, characterized by $\xi $.
However, this dependence is very strong only for $\xi \ll 1$ when the second
term inside the bracket in (\ref{FI7}) dominates. For $\xi =0.6$ we get $%
l_{S}\simeq 1100$ if $\Omega _{m}h_{75}^{2}\simeq 0.3$ and $l_{S}\simeq 980$
for $\Omega _{m}h_{75}^{2}\simeq 1.$ The dissipation scale in the universe
with more cold matter is bigger (correspondingly $l_{S}$ is smaller) because
in this case the recombination happens at later cosmic time $t_{r}$ and
hence the perturbations get an extra time to be washed out.

\textit{The parameter} $\rho ,$ which determines the location of the peaks,
can be easily calculated if one substitutes the speed of sound 
\begin{equation}
c_{s}\left( \eta \right) =\frac{1}{\sqrt{3}}\left( 1+\xi \left( \frac{a(\eta
)}{a\left( \eta _{r}\right) }\right) \right) ^{-1/2}  \label{Fss}
\end{equation}%
where $a(\eta )$ is given by (\ref{dustradsol}), into (\ref{Fshad}) and
performs an explicit integration there. The result is 
\begin{equation}
\varrho \simeq \frac{I_{\Lambda }}{\sqrt{3z_{r}\xi }}\ln \left( \frac{\sqrt{%
\left( 1+z_{r}/z_{eq}\right) \xi }+\sqrt{\left( 1+\xi \right) }}{1+\sqrt{\xi
\left( z_{r}/z_{eq}\right) }}\right)  \label{Fro}
\end{equation}%
It is clear that $\varrho $ depends on both baryon and matter densities.
However, it is not very transparent how $\varrho $ behaves when we change
these parameters. Therefore it is worthwhile to find a simple numerical fit
for (\ref{Fro}), which would reproduce the parameter dependence of $\varrho $
within reasonable range of change of $\xi $ and $\Omega _{m}h_{75}^{2}.$ The
fit\footnote{%
I am very grateful to P. Steinhardt for helping me to check numerically the
accuracy of this fit and the fits (\ref{Fflt3})-(\ref{Fflt5}).} 
\begin{equation}
\varrho \simeq 0.014\left( 1+0.13\xi \right) ^{-1}\left( \Omega
_{m}h_{75}^{2}\right) ^{1/4}I_{\Lambda }  \label{Fro1}
\end{equation}%
reproduces the exact result (\ref{Fro}) with the accuracy about $5\div 7\%$
or better everywhere in the region $0<\xi <5,$ $0.1<\Omega _{m}h_{75}<1$
where the function $\varrho $ itself changes in about three times. Combining
this with the numerical fit for $I_{\Lambda }$ in (\ref{FI1}) we have 
\begin{equation}
\varrho \simeq 0.014\left( 1+0.13\xi \right) ^{-1}\left( \Omega
_{m}h_{75}^{3.1}\right) ^{0.16}  \label{Fro2}
\end{equation}

\ 

\bigskip

\textit{The transfer functions} $T_{p},T_{o}$ depend only on $k\eta _{eq}$
and can be expressed as the functions of variable $x=k\eta _{0}/l:$%
\begin{equation}
k\eta _{eq}=\frac{\eta _{eq}}{\eta _{0}}lx\simeq 0.72\left( \Omega
_{m}h_{75}^{2}\right) ^{-1/2}I_{\Lambda }l_{200}x  \label{Fkn}
\end{equation}%
where $l_{200}\equiv l/200.$ As we will see the contributions to the
integrals defining the fluctuations in the region of the first few acoustic
peaks comes from $O\left( 1\right) >x\geq 1.$ Therefore for $200<l<1000$ the
transfer functions in the relevant range of $k\eta _{eq}$ can be
approximated by (\ref{Ftf1}), (\ref{Ftf2}); hence 
\begin{equation}
T_{p}\left( x\right) =0.74-0.25\left( P+\ln x\right)  \label{Ftp1}
\end{equation}%
where 
\begin{equation}
P\left( l,\Omega _{m},h_{75}\right) \equiv \ln \left( \frac{I_{\Lambda
}l_{200}}{\sqrt{\Omega _{m}h_{75}^{2}}}\right) ,  \label{Fdp}
\end{equation}%
and, respectively,%
\begin{equation}
T_{o}\left( x\right) =0.5+0.36\left( P+\ln x\right)  \label{Fdo1}
\end{equation}%
\ 

\subsection{Calculating the spectrum (continuation)}

Now I will proceed with the calculations of the fluctuations. The main
contribution to the integrals from the oscillating functions (\ref{Fclo1}), (%
\ref{Fclo2}) gives the vicinity of the singular point $x=1.$ These integrals
have the form%
\begin{equation}
\int_{1}^{\infty }\frac{f\left( x\right) \cos \left( ax\right) }{\sqrt{x-1}}%
dx  \label{w}
\end{equation}%
and after making substitution $x=y^{2}+1$ can be calculated using stationary
(saddle) point method. The result is 
\begin{equation}
\int_{1}^{\infty }\frac{f\left( x\right) \cos \left( ax\right) }{\sqrt{x-1}}%
dx\approx \frac{f\left( 1\right) }{\left( 1+B^{2}\right) ^{1/4}}\sqrt{\frac{%
\pi }{a}}\cos \left( a+\frac{\pi }{4}+\frac{1}{2}\arcsin \frac{D}{\sqrt{%
1+D^{2}}}\right) ,  \label{cor}
\end{equation}%
where $D\equiv \left( d\ln f/adx\right) _{x=1}.$ For big $a$ we can put $%
D\approx 0$ and the above formula simplifies to%
\begin{equation}
\int_{1}^{\infty }\frac{f\left( x\right) \cos \left( ax\right) }{\sqrt{x-1}}%
dx\approx f\left( 1\right) \sqrt{\frac{\pi }{a}}\cos \left( a+\frac{\pi }{4}%
\right)  \label{Fsim}
\end{equation}

\bigskip Using (\ref{Fsim}) to calculate the integrals in (\ref{Fclo1}), (%
\ref{Fclo2}) we obtain:%
\begin{equation}
O\simeq \sqrt{\frac{\pi }{\varrho l}}\left( \mathcal{A}_{1}\cos \left(
l\varrho +\pi /4\right) +\mathcal{A}_{2}\cos \left( 2l\varrho +\pi /4\right)
\right)  \label{Fflt00}
\end{equation}%
where the coefficients 
\begin{equation}
\mathcal{A}_{1}\equiv -\left( \frac{4}{3\left( 1+\xi \right) }\right)
^{1/4}\xi \left( T_{p}T_{o}\right) _{x=1}e^{\left( \frac{1}{2}\left(
l_{S}^{-2}-l_{f}^{-2}\right) l^{2}\right) },\mbox{ \ \ }\mathcal{A}%
_{2}\equiv \frac{\left( T_{o}^{2}\right) _{x=1}}{4\sqrt{3\left( 1+\xi
\right) }}  \label{Fflt0}
\end{equation}%
are slowly varying functions of $l.$ They also depend on the basic
cosmological parameters and the spectrum of the fluctuations at $l>200$ is
rather sensitive to the variation of these parameters. It is worth to
mention that in this approximation the contribution of the Doppler term to
the oscillating part of the spectrum drops out. One can check that actually
this contribution at $l>200$ do not exceed few percent of the total
amplitude. If $\Omega _{b}\ll \Omega _{m}$ the transfer functions for the
most interesting range $200<l<1000$ can be approximated by (\ref{Ftp1}) and (%
\ref{Fdo1}). In this case we have 
\begin{equation}
\mbox{\ \ }\mathcal{A}_{1}\simeq 0.1\frac{\left( \left( P-0.78\right)
^{2}-4.3\right) \xi }{\left( 1+\xi \right) ^{1/4}}e^{\left( \frac{1}{2}%
\left( l_{S}^{-2}-l_{f}^{-2}\right) l^{2}\right) },\mbox{ \ \ }\mathcal{A}%
_{2}\simeq 0.14\frac{\left( 0.5+0.36P\right) ^{2}}{\left( 1+\xi \right)
^{1/2}}  \label{Fflt2}
\end{equation}%
$\allowbreak $ where $P$ is given by (\ref{Fdp})

Substituting (\ref{Ftp1}) in the expression (\ref{FN1}) for non-oscillating
contribution $N_{1}$ we get%
\begin{equation}
N_{1}\simeq \xi ^{2}\left[ \left( 0.74-0.25P\right) ^{2}I_{0}-\left(
0.37-0.125P\right) I_{1}+\left( 0.25\right) ^{2}I_{2}\right]  \label{FN1a}
\end{equation}%
where the integrals%
\begin{equation}
I_{m}\left( l/l_{f}\right) \equiv \int_{1}^{\infty }\frac{\left( \ln
x\right) ^{m}}{x^{2}\sqrt{x^{2}-1}}e^{-\left( l/l_{f}\right) ^{2}x^{2}}dx
\label{FNi}
\end{equation}%
can be calculated in terms of the hypergeometric functions. However the
obtained expressions are not very transparent and therefore it makes sense
to find the numerical fits for them. The final result is 
\begin{equation}
N_{1}\simeq 0.063\xi ^{2}\frac{\left( P-0.22\left( l/l_{f}\right)
^{0.3}-2.6\right) ^{2}}{1+0.65\left( l/l_{f}\right) ^{1.4}}e^{-\left(
l/l_{f}\right) ^{2}}.  \label{Fflt3}
\end{equation}%
Similar by, we obtain 
\begin{equation}
N_{2}\simeq \frac{0.037}{\left( 1+\xi \right) ^{1/2}}\frac{\left(
P-0.22\left( l/l_{S}\right) ^{0.3}+1.7\right) ^{2}}{1+0.65\left(
l/l_{S}\right) ^{1.4}}e^{-\left( l/l_{S}\right) ^{2}}  \label{Fflt4}
\end{equation}%
The Doppler contribution to nonoscillating part of the spectrum is
comparable to $N_{2}$ and is equal to 
\begin{equation}
N_{3}\simeq \frac{0.033}{\left( 1+\xi \right) ^{3/2}}\frac{\left(
P-0.5\left( l/l_{S}\right) ^{0.55}+2.2\right) ^{2}}{1+2\left( l/l_{S}\right)
^{2}}e^{-\left( l/l_{S}\right) ^{2}}  \label{Fflt5}
\end{equation}%
The numerical fits (\ref{Fflt3})-(\ref{Fflt5}) reproduce the exact result in
the most interesting range of multipoles with a few percent accuracy for a
wide range of cosmological parameters. The extra dependence on $l/l_{S}$ and 
$l/l_{f}$ is due to the fact that the exponent in the integrals from
nonoscillating functions can not be just simply estimated at $x=1.$ When the
expression under the integral is monotonic function the substantial
contribution to it comes not only from the vicinity of $x=1$ but also from $%
x\sim O\left( 1\right) .$ The nonoscillating contribution of the Doppler
term given by $N_{3}$ is rather essential and can not be ignored.

It is convenient to normalize $l\left( l+1\right) C_{l}$ for big $l$ to the
amplitude of fluctuations for small $l,$ given by (\ref{Fsp}), so that
finally we obtain%
\begin{equation}
\frac{l\left( l+1\right) C_{l}}{\left( l\left( l+1\right) C_{l}\right)
_{l<30}}=\frac{100}{9}\left( O+N_{1}+N_{2}+N_{3}\right) .  \label{Fflt6}
\end{equation}%
where $O,N_{1},N_{2},N_{3}$ are respectively given by (\ref{Fflt00}), (\ref%
{Fflt3}), (\ref{Fflt4}), (\ref{Fflt5}). In the case of the concordance model
($\Omega _{m}=0.3,\Omega _{\Lambda }=0.7,\Omega _{b}=0.04$ and $%
H=70km/sec\cdot Mpc$) the result is presented in Fig.1, where I have
separately shown by the dashed and thin solid lines, respectively, the
overall nonoscillating and oscillating contributions. The total resulting
fluctuations are shown by the thick solid line.

\begin{figure}[h]
\begin{center}
\includegraphics[width=11cm,angle=0]{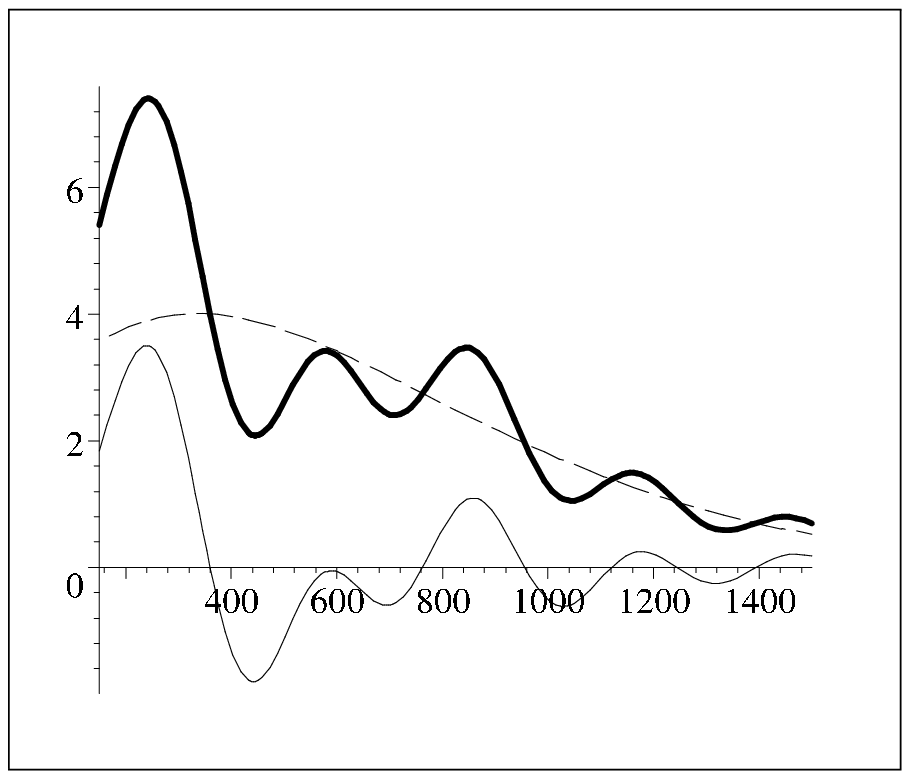} \vspace{0.1cm}\\[0pt]
figure 1
\end{center}
\end{figure}

\textit{About accuracy}\textbf{.} Comparing (\ref{Fflt6}) to CMBFAST runs%
\footnote{%
I am very grateful to P. Steinhardt, S. Bashinsky and U.Seljak for
performing numerous CMBFAST runs necessary for this work.} one can easily
check that the analytical approximation works rather well reproducing the
numerical results with good accuracy in a rather wide range of the
cosmological parameters around concordance model. Namely, for $\Omega
_{m}=0.3$ the agreement is still very good up to $\Omega _{b}\simeq 0.08,$
when the baryons constitute already about $30\%$ percent of the total cold
matter density. At higher $\Omega _{b}$ the contribution of the baryons to
the gravitational potential which we neglected becomes very essential and
one can not use anymore the analytical formula (\ref{Fflt6}). This formula
was derived under assumption $\Omega _{b}\ll \Omega _{m}$ and is not
trustable when the baryons constitute a very substantial fraction of the
total amount of cold matter. It is also worth to mention that at high $%
\Omega _{m}h_{75}^{2}$ the expected accuracy in the region of the first peak
is not as good as in the region of the second and third peaks . This is
because the used approximations for transfer functions become not so
accurate on the border of the interval corresponding to $1<k\eta _{eq}<10$.
In particular, if $\Omega _{m}h_{75}^{2}=1$ the main contribution to the
first acoustic peak located at $l\sim 200$ give the perturbations with the
wavenumbers $k\eta _{eq}\sim 0.7$ (see (\ref{Fkn})) where the approximations
(\ref{Ftf1}) and (\ref{Ftf2}) are not very accurate. Hence, although for the
model with $\Omega _{m}h_{75}^{2}=1$ and $\Omega _{b}=0.04$ the analytical
result is still in a fair agreement with the numerics, its accuracy in the
region of the first peak is not as good as for concordance model. Also note
that the peaks given by (\ref{Fflt6}) are shifted by about $10\%$ compared
to the numerical results. One of the reasons for that is that with the
purpose to simplify the final expression we neglected in (\ref{cor}) an
extra phase shift proportional to $D.$ The other reason is that we
underestimated the parameter $\varrho $ which was derived in the assumption
of instantaneous recombination. In reality the recombination takes place
within about quarter of the cosmological time and in the process of
recombination the baryons decouple from the radiation. As a result the sound
speed increases and $\varrho $ should be a bit bigger compared to (\ref{Fro}%
).

However the main value of the analytical result is not in its competitive
accuracy with the numerics, but because it allows us to understand the main
features of the CMB spectrum and study explicitly how they depend on the
cosmological parameters. In turn it opens a possibility to understand the
degeneracy of the spectrum with respect to the certain combinations of the
cosmological parameters which could lead to a \textquotedblright cosmic
confusion\textquotedblright .

\section{Determining the cosmological parameters}

Let us discuss how the main features of the spectrum change when the
cosmological parameters vary. These parameters are: the amplitude $B$ and
the slope $n_{s}$ of the primordial spectrum, the baryon density
characterized by $\Omega _{b}$, the total cold matter density $\Omega _{m},$
the cosmological constant $\Omega _{\Lambda }$ and the Hubble constant $%
h_{75}.$ The amplitude and the slope of the primordial spectrum can already
be determined with a reasonably good accuracy when we consider only the
measurements in big angular scales. From these observations it follows that
the spectrum does not deviate too much from the scale invariant\ ($n_{s}=1$%
). The cosmic variance, which is important in the big scales which are not
much disturbed by the transfer functions does not allow us to conclude
anything about small deviations from scale invariant spectrum predicted by
inflation on basis of only these observations.

The results for the fluctuations in small angular scales were derived
assuming a flat universe, where $\Omega _{m}+\Omega _{\Lambda }=1,$ and a
scale invariant spectrum with $n_{s}=1.$ How do they change when the
spectrum deviates from the scale invariant will be pointed out below. First,
I would like to concentrate on flat models with scale invariant spectrum ($%
n_{s}=1$) and find how the characteristic features of the CMB spectrum
depend on the cosmological parameters $\Omega _{b},\Omega _{m}$ and $h_{75}$
(the cosmological constant is fixed by the flatness condition to be $\Omega
_{\Lambda }=1-\Omega _{m}$).

\textit{The location of the peaks and the flatness of the universe.} The
most interesting feature of the spectrum is the presence of the peaks and
valleys, the height and location of which very sensitively depend of the
major cosmological parameters. At $l>1000$ the fluctuations are strongly
suppressed and therefore the most interesting part of the spectrum is those
one where the first three peaks are located. These peaks arise as a result
of superimposing of the oscillating contribution to the fluctuations $O$,
given by (\ref{Fflt00}), on the \textquotedblright hill\textquotedblright\ $%
N\left( l\right) =N_{1}+N_{2}+N_{3}$ representing a nonoscillating part of
the spectrum (see Fig.1). It is clear that the locations and the heights of
the peaks depend not only on the oscillating \ part, but also on the shape
of the \textquotedblright hill\textquotedblright . Let us neglect for a
moment the effect of the \textquotedblright hill shape\textquotedblright .
In this case the location of the peaks would be determined by the
superposition of two cosines in (\ref{Fflt00}). If $\left\vert \mathcal{A}%
_{1}\right\vert \ll \mathcal{A}_{2}$ the peaks should be located at%
\begin{equation}
l_{n}=\pi \varrho ^{-1}\left( n-\frac{1}{8}\right)   \label{Fm}
\end{equation}%
where $n=1,2,3...$ and $\varrho $ is given by (\ref{Fro1}). The first term
in (\ref{Fflt00}) has twice bigger period than the second and its amplitude $%
\mathcal{A}_{1}$ is negative. Therefore it participates in the constructive
interference for the odd peaks ($n=1,3,...$) and in destructive interference
for the even peaks ($n=2,4,...$). Moreover, because of the shift of the
arguments of two cosines, the maxima of these cosines do not coincide and,
as a result, first and third peaks (for which the interference is
constructive) should be located in between the appropriate maxima of these
two cosines, that is, at%
\begin{equation}
l_{1}\simeq \left( \frac{6}{8}\div \frac{7}{8}\right) \pi \varrho ^{-1},%
\mbox{ }l_{3}\simeq \left( 2\frac{6}{8}\div 2\frac{7}{8}\right) \pi \varrho
^{-1},  \label{Fm1}
\end{equation}%
where the symbol $\div $ denotes the appropriate interval. If $\left\vert 
\mathcal{A}_{1}\right\vert \gg \mathcal{A}_{2}$ the peaks move closer to the
lower bounds of the intervals in (\ref{Fm1}). In fact, the situation is more
complicated because the nonoscillating contribution $N$ is not constant but
is represented by \textquotedblright hill\textquotedblright . As it is clear
from Fig.1, this leads to the further shift of the peaks to the
\textquotedblleft top of the hill". For instance, for concordance model
first peak moves a bit to the right, while the third peak to the left.
Substituting $\xi \simeq 0.6$ and $\Omega _{m}h_{75}^{2}\simeq 0.26$ into (%
\ref{Fro}) we find that for this model the first peak should be located at $%
l_{1}\simeq 225\div 265,$ that is, somewhere in between $225$ and $265.$ For
the third peak $l_{3}\simeq 825\div 865.$ Because of the reasons I mentioned
above, this result should be corrected by about $10\%$ shifting the first
peaks to the left.

In the region of the odd peaks one has destructive interference of the
oscillating terms. The first term in (\ref{Fflt00}) which takes the minimal
(negative) value tries to annihilate these peaks. The second peak (if it
exists), should be located at%
\begin{equation}
l_{2}\simeq \left( 2\frac{6}{8}\div 2\frac{7}{8}\right) \pi \varrho ^{-1}
\label{Fm2}
\end{equation}%
or in the concordance model at $l_{2}\simeq 525\div 565$.

How sensitive is the peaks location to the variation of the cosmological
parameters? According to (\ref{Fro}) (see also (\ref{Fro2})) $\varrho $
changes when the baryon and cold matter densities vary and therefore (see (%
\ref{Fm1}), (\ref{Fm2})) the peaks location should also depend on these
parameters. The parameter $\varrho $ is not very sensitive function of $%
\Omega _{m},h_{75}$ and $\xi $. Therefore, the location of the first peak in
a flat universe is relatively stable when we vary these parameters. In
particular, when the baryon density increases in two times ($\xi \simeq
0.6\rightarrow $ $\xi \simeq 1.2$) the first peak moves to the right by $%
\Delta l_{1}\sim +20$ and the shift of the second and third peaks are,
respectively, $\Delta l_{2}\sim +40$ and $\Delta l_{3}\sim +60.$ When
determining the location of the peaks, the baryon density always enters in
combination $\xi \propto \Omega _{b}h_{75}^{2}$ with the Hubble constant.
The cold matter density comes together with $h_{75}$ as $\Omega
_{m}h_{75}^{3.1}.$ The increase of the cold matter density has an effect
opposite to the increase of the baryon density, namely, if for given $\xi
\simeq 0.6$ the cold matter density increases twice ($\Omega
_{m}h_{75}^{3.1}\simeq 0.3\rightarrow $ $\Omega _{m}h_{75}^{3.1}\simeq 0.6$%
), the first peak goes to the left by $\Delta l_{1}\sim -20$ and
respectively $\Delta l_{2}\sim -40$ and $\Delta l_{3}\sim -60.$ Thus we see
that even in a flat universe, one can shift the location of the first peak
quite substantially ($\Delta l_{1}\sim 40$) increasing the baryon density
twice and simultaneously decreasing the cold matter density by the same
factor.

Why in this case can we be sure that the first peak location is a good
indicator of the universe curvature? Fortunately, if we will fix the height
of the first peak, then its location becomes \textquotedblleft stable" with
respect to the admitted variations of the cosmological parameters. The
height of the first peak sensitively depends of the cold matter and baryon
density. Given the height of this peak we can still vary the baryon and cold
matter densities together. However if the cold matter density would increase
and we would like still to keep the height of the peak to be the same, we
have simultaneously change the baryon density, namely, it also should
increase. Since the change of the baryon and cold matter densities have
opposite effects on the peak location, it will be shifted not very much if
both of them will increase simultaneously. For instance, if both densities
increase by a factor two around concordance model, one can expect that $%
\Delta l_{1}\sim 0.$ This explains the stability of the location of the
first peak for the acceptable range of change of the cosmological parameters
in a flat universe. The obtained result on the location of the first Doppler
peak and its relative stability to the variation of the unknown cosmological
parameters is a fair agreement with numerical calculations. The stability of
the first peak location makes it an irreplaceable indicator of the total
energy density of the universe. Actually, the peak location is incomparably
more sensitive to the total energy density (in the open universe without
cosmological constant $l_{1}\propto \Omega _{tot}^{-1/2}$). The present
observations strongly favor a flat universe ($\Omega _{tot}=1$), as
predicted by inflation.

\textit{Height of the peaks and the baryon and cold matter densities.} In
concordance model the amplitude of the first acoustic peak is in about $%
7\div 8$ times bigger than the amplitude of the fluctuations in big angular
scales. Substituting $l_{n},$ given by (\ref{Fm1}), (\ref{Fm2}), into (\ref%
{Fdp}) and using the formula (\ref{Fro}) for $\varrho $ we see that the
factor $I_{\Lambda }$ is cancelled in the expression for $P$ and therefore
the height of the peaks given by (\ref{Fflt6}) estimated at $l_{n}$ can
depend only on $\Omega _{m}h_{75}^{2}$ and $\Omega _{b}h_{75}^{2}$ (or $\xi $%
). If, for fixed $\Omega _{m}h_{75}^{2},$ one increases the baryon density,
the height $H_{1}\left( \Omega _{m}h_{75}^{2},\xi \right) $ also increases.
In the concordance model the increase of the baryon density by factor two
(from $\xi \simeq 0.6$ to $\xi \simeq 1.2$) leads to the increase of the
amplitude $H_{1}$ in $1,5$ times. This increase in the amplitude is mostly
due to two terms in (\ref{Fflt6}), $N_{1}$ (proportional to $\xi ^{2}$) and $%
O$ (since $\mathcal{A}_{1}\propto \xi $). In turn, the increase of the cold
matter density (at fixed $\xi $) suppresses the height of the first peak $%
H_{1}.$ It becomes clear why this happens if we note that for fixed $l,$ the
function $P$ entering the formulae for fluctuations decreases when $\Omega
_{m}h_{75}^{2}$ increases. As a result the overall amplitude of the first
peak decreases (mainly because $N_{2}$ and $N_{3}$ contributions decrease
when $\Omega _{m}h_{75}^{2}$ increases). Therefore the height of the first
peak is degenerate with respect to a certain combination of the baryon and
cold matter densities. In a certain range of parameters the increase of the
height due to the baryon density can be compensated if we simultaneously
increase the cold matter density. However, if the baryon density would be
too high, the increase of the height of first peak could not be anymore
compensated by increase in $\Omega _{m}h_{75}^{2}$ because $\Omega
_{m}h_{75}^{2}$ can not much exceed unity. (Moreover, for big $\Omega
_{m}h_{75}^{2}$ the transfer functions responsible for $\Omega
_{m}h_{75}^{2}-$dependence of $H_{1}$ reach their asymptotic values for
those values of $k\eta _{eq}$ which mainly contribute to the fluctuations in
the region of the first peak). \textit{Hence, just relying on the result
about the height of the first peak one can safely conclude that the baryon
density can not be more than }$15\div 20\%$\textit{\ of the total critical
density}.

The degeneracy in determining $\Omega _{m}h_{75}^{2}$ and $\xi $ parameters
can be easily resolved if we consider the second peak, which results mostly
from the destructive interference of the oscillating terms in (\ref{Fflt00})
superimposed on the \textquotedblright hill\textquotedblright\ given by $N$%
-contribution. In the concordance model this peak is strongly suppressed in $%
O-$contribution and partially recovered only in the resulting spectrum
because of the $N-$contribution (as one can see in Fig.1 the
\textquotedblright hill\textquotedblright\ has a sufficiently steep decline
in this region). The presence of the second peak sensitively depends on the
ratio of the amplitudes $\mathcal{A}_{1}$ and $\mathcal{A}_{2}.$ Since the
amplitude $\mathcal{A}_{1}$ of the first term in (\ref{Fflt00}), which tries
to \textquotedblright kill\textquotedblright\ the peak is proportional to
the baryon density $\xi $, while $\mathcal{A}_{2}$ slightly decreases when $%
\xi $ grows, one can expect that the presence of large amount of baryons
should diminish and may be even completely remove the second peak. Actually
the \textquotedblleft $O-$contribution" to the peak disappears when the
baryon density increases only twice compared to its value in concordance
model. However, in the resulting spectrum this peak still survives. This is
because the growing amount of baryons simultaneously amplifies $N_{1}-$%
contribution to nonoscillating part of the spectrum and in turn this
significantly steepens the \textquotedblleft hill" in the region where the
second peak is located. The analytical formulae become inapplicable at very
high baryon densities. However the numerical calculations show that for $%
\Omega _{m}h_{75}^{2}\simeq 0.26$ the second peak is still present and has
nearly the same amplitude as the third peak even if baryons constitute about 
$70\%$ of all cold matter. Hence the presence of the second peak can not 
\textit{alone} be considered as the indication of the low baryon density.
Nevertheless, \textit{in combination with the observed height of the first
peak the second peak} \textit{is a very sensitive indicator not only for the
baryon density, but also for total cold matter density.} Given the height of
the first peak, we can still vary the baryon and cold matter densities
increasing or decreasing them simultaneously, since they \textquotedblleft
act in opposite directions". However they influence differently the second
peak. Namely, the simultaneous increase of the baryon and cold matter
densities tries to \textquotedblleft annihilate this peak". Actually, the
amplitude of the second peak depends on the amplitudes $\mathcal{A}%
_{1}\left( \xi ,\Omega _{m}h_{75}^{2}\right) $ and $\mathcal{A}_{2}\left(
\xi ,\Omega _{m}h_{75}^{2}\right) $ in superposition of two cosines in (\ref%
{Fflt00}). The increase of the baryon density tends to \textquotedblright
kill\textquotedblright\ this second acoustic peak. The increase of the cold
matter density at fixed $\xi $ has a similar effect. This is because $%
\mathcal{A}_{2}\propto \left( T_{o}^{2}\right) _{x=1}$ decreases faster than 
$\mathcal{A}_{1}\propto \left( T_{o}T_{p}\right) _{x=1}$ when $\Omega
_{m}h_{75}^{2}$ increases. At big $\Omega _{m}h_{75}^{2}$ the term which
\textquotedblleft kills the peak" dominates. Hence the height of the second
peak depends simultaneously on the baryon and total cold matter densities
and is very sensitive to the independent variation of both of them. Fixing
the relation between $\xi $ and $\Omega _{m}h_{75}^{2}$ from the height of
the first peak we can find the particular values of these parameters
measuring the height of the second acoustic peak. For instance, if $\Omega
_{m}h_{75}^{2}=1,$ then $\xi $ should be about unity if one want to get the
height of the first peak to be in agreement with observations. In this case
second peak completely disappears. Hence \textit{the experimental detection
of the second peak proves that the total density of the cold matter is
smaller that the critical one and the baryon density is smaller than }$6\div
8\%.$This is in an excellent agreement with nucleosynthesis bounds. Shortly
both of the results could be formulated as \textquotedblright too much
baryons would destroy all deuterium and kill the second acoustic
peak\textquotedblright . In combination with the location and height of the
first peak the presence of the second peak is also a strong independent
indicator of the dark energy in the universe. In fact, \textit{from the
location of the first peak it follows that the total density in the universe
is critical and the presence of the second peak means that the cold matter
can constitute only the fraction of it}.

Since the heights and locations of the peaks depend on the different
combination of $\Omega _{m}$ and $h_{75}$ this allows us to resolve the
degeneracy in determining the Hubble constant. As we have seen for a given $%
\Omega _{b}h_{75}^{2}$ the location of the peaks depends on $\Omega
_{m}h_{75}^{3.1},$ while their heights is determined by $\Omega
_{m}h_{75}^{2}.$ Therefore keeping $\Omega _{b}h_{75}^{2}$ and $\Omega
_{m}h_{75}^{2}$ to be fixed by the heights of the peaks we can still vary
the Hubble parameter $h_{75}$ shifting the position of the peaks. As it
follows from (\ref{Fm1}), (\ref{Fro2}), for the given $\Omega
_{b}h_{75}^{2}\simeq 0.04$ and $\Omega _{m}h_{75}^{2}\simeq 0.3$ the
increase of the Hubble constant by $20\%$ (say from $70$ $km/sec\cdot Mpc$
to $85$ $km/sec\cdot Mpc$) moves the peaks to the left by $3\%,$ that is, $%
\Delta l_{1}\simeq 7$ and $\Delta l_{2}\simeq 15.$ Hence if we want to get
an accurate determination of the Hubble constant from CMB spectrum \textit{%
alone} we have to know the location of the peaks with very high accuracy. If
the location of the peaks will be determined with $1\%$ accuracy then the
expected accuracy of the Hubble constant will be about $7\%.$

Up to now we were assuming that the primordial spectrum of the
inhomogeneities is scale invariant, that is the spectral index is $n_{s}=1.$
The inflation predicts that there should be deviations from the scale
invariant spectrum and we expect that $n_{s}\simeq 0.92\div 0.97.$ The above
derivation for the CMB fluctuations can be easily modified to account for
these deviations.

If $n_{s}\neq 1$ the obtained amplitudes of the fluctuations at given $l$
should be just multiplied by the factor proportional to $l^{1-n}.$ To
resolve the degeneracy in determining the cosmological parameters in this
case the heights and location of the first two peaks are not sufficient.
Actually for a given $n$ one can always find the combination of the $\Omega
_{b}h_{75}^{2}$ and $\Omega _{m}h_{75}^{2}-$ parameters to fit the heights
of the first two peaks. The location of these peaks is also not very
sensitive to the deviations of the spectral index from unity. Therefore one
needs extra information. With this purpose we can use for instance the
height of the third acoustic peak. As one can check the height of this peak
is not so sensitive to $\Omega _{b}h_{75}^{2}$ and $\Omega _{m}h_{75}^{2}$
as for the first two peaks. Fixing these parameters and varying the spectral
index $n_{s}$ for a given unchanged height of the first peak (this can
always be done if together with $n_{s}$ we vary the amplitude of the
spectrum $B$) we find that the relative height of the third peak changes as%
\begin{equation}
\frac{\Delta H_{3}}{H_{3}}\sim \left( \frac{l_{3}}{l_{1}}\right) ^{1-n_{s}}-1
\label{F3}
\end{equation}%
For instance, if $n_{s}\simeq 0.95$ the height of the third peak increases
by about $5\%$ compared to the case of $n_{s}=1.$ From this estimate one can
get a rough idea about necessary accuracy of the measurements to find the
expected deviations from the scale invariant spectrum.

\section{\protect\bigskip Conclusions}

In this paper I have shown that if we assume that the main ingredients of
the cosmological model are known, then we can completely resolve the
degeneracy and determine the main cosmological parameters from the CMB
spectrum. For that we just need to know the main features of the spectrum,
namely, the heights and location of the peaks. Of course, the accuracy of
the determination is different for different parameters and seems to be the
worst for the Hubble constant. The information we gain in the observations
exceeds the discussed features of the spectrum. Namely, one measures also
the entire shape of the spectrum, which, of course, also depends on the
cosmological parameters. The necessity to fit this shape restricts the
possible values of the parameters even in the case when we have the
measurements only in the region of the first peak. This shape (as well as
the heights and location of the peaks) also depends on the dissipation
scales $l_{f}$ and $l_{S},$ which in turn slightly depends on the
cosmological parameters. For the concordance model $l_{S}\sim 1000$, and it
is clear that the dissipation does not influence very much the first peak
and becomes very essential in the region of the second peak and at high $l.$
In particular, at $l>1000$ this effect entirely dominates, leading to the
exponential falloff of the spectrum at very high multipoles. This falloff is
very sensitive to the parameters and, being measured, can give us extra
information about them. The measurements of the polarization provides
additional valuable information about the cosmological parameters. When we
vary the parameters the detailed behavior of the spectrum is, of course,
more complicated than I described above (I also neglected here the
primordial gravity waves which can give rather substantial contribution at $%
l<30$). However, the above consideration correctly reflects the main
features of this behavior and gives the physical understanding why the CMB
spectrum so sensitively depends on the cosmological parameters.

\section*{Acknowledgments}

This work was done, in part, during my sabbatical stay at Princeton
University. It is the pleasure for me to thank Physics Department for warm
hospitality. I am very grateful to U. Seljak for many hours clarifying
discussions on CMB fluctuations without which I probably would not be able
to proceed. My special thanks to P. Steinhardt for invitation to spend
sabbatical in Princeton, his hospitality, encouraging remarks, discussions
and help with numerics. I would also like to thank S. Bashinsky, D. Bond, L.
Page and S. Weinberg for discussions and illuminating remarks which were
very helpful.

\appendix

\section{Hydrogen recombination}

The equilibrium description of recombination by Saha's formula fails nearly
immediately after the beginning of recombination when only a few percent of
hydrogen becomes neutral. Therefore one has to use the kinetic approach to
describe the noninstantaneous (delayed) recombination\cite{SKZ}.

The direct recombination to the ground state with the emission of one
energetic photon is not very efficient. The emitted photon has enough energy
to immediately ionize the first neutral hydrogen atom it meets. One can
easily check that the two competing processes, direct recombination to the
ground state and ionization, occur with a very high rate leaving no net
contribution. More efficient is the cascade recombination when the neutral
hydrogen is first formed in the excited state and then goes to the ground
state. However, even in the cascade recombination at least one very
energetic photon is emitted. Its energy corresponds to the energy difference
between $2p-$ and $1s-$states. This Lyman-alpha photon $\left( L_{\alpha
}\right) $ has the energy $3B_{H}/4\Rightarrow 117000%
{{}^\circ}%
K$ and a rather big resonance absorption cross-section which at the
recombination temperature is about $\sigma _{\alpha }\simeq 10^{-17}\div
10^{-16}$ $cm^{2}.$ The $L_{\alpha }-$photons are reabsorbed in $\tau
_{\alpha }\simeq \left( \sigma _{\alpha }n_{H}\right) ^{-1}\sim 10^{3}\div
10^{4}$ $sec$ after emission. This time has to be compared to the
cosmological time. During the matter dominated epoch the cosmological time
can be easily expressed through the temperature if we just equate the energy
density of the cold particles to the critical energy density $\varepsilon
^{cr}=1/\left( 6\pi t^{2}\right) $ and note that $T=T_{\gamma 0}\left(
1+z\right) ;$ hence%
\begin{equation}
t_{sec}\simeq 2.75\times 10^{17}\left( \Omega _{m}h_{75}^{2}\right)
^{-1/2}\left( \frac{T_{\gamma 0}}{T}\right) ^{3/2},  \label{ct}
\end{equation}%
where $h_{75}$ is the Hubble constant normalized on $75$ $km/sec\cdot Mpc,$ $%
T_{\gamma 0}\simeq 2.72$ $K$ and $\Omega _{m}$ is the contribution of cold
matter to the critical density. At the moment of recombination $\tau
_{\alpha }\ll t_{c}\sim 10^{13}$ $sec$ and the $L_{\alpha }-$quanta are not
significantly redshifted before they are reabsorbed. Therefore below I will
neglect the redshifts of these quanta, which could in principle take them
outside of the resonance line. The presence of the big number of $L_{\alpha
} $ photons leads to an overabundance of the electrons $\left( e\right) $,
protons $\left( p\right) $ and $2s,p-$ states of the neutral hydrogen,
compared to what is predicted by the equilibrium Saha's formula. In turn,
this delays the recombination and for a given temperature the actual degree
of ionization exceeds its equilibrium value given by the equilibrium Saha's
formula. The full system of kinetic equations describing the recombination
is rather complicated and can be solved only numerically. Here I will
present the useful approximate treatment, which is in a very good agreement
with the results of the numerical calculations.

I neglect all highly excited hydrogen states so that the main
\textquotedblright players left in the game\textquotedblright\ will be $1s,$ 
$2s,$ $2p-$states together with the electrons, protons, thermal photons, $%
L_{\alpha }$- and other energetic quanta emitted during recombination. The
main processes, involving these components, are symbolically shown in Fig.2.
They are responsible for the \textquotedblleft converting elements" and
bring them from one \textquotedblleft reservoir" to the other changing the
appropriate concentrations. First note that the recombination acts directly
to the ground state do not lead to any net change in the system and leave
the concentrations in the appropriate \textquotedblleft reservoirs" in Fig.2
without change.

\begin{figure}[h]
\begin{center}
\includegraphics[width=9cm,angle=0]{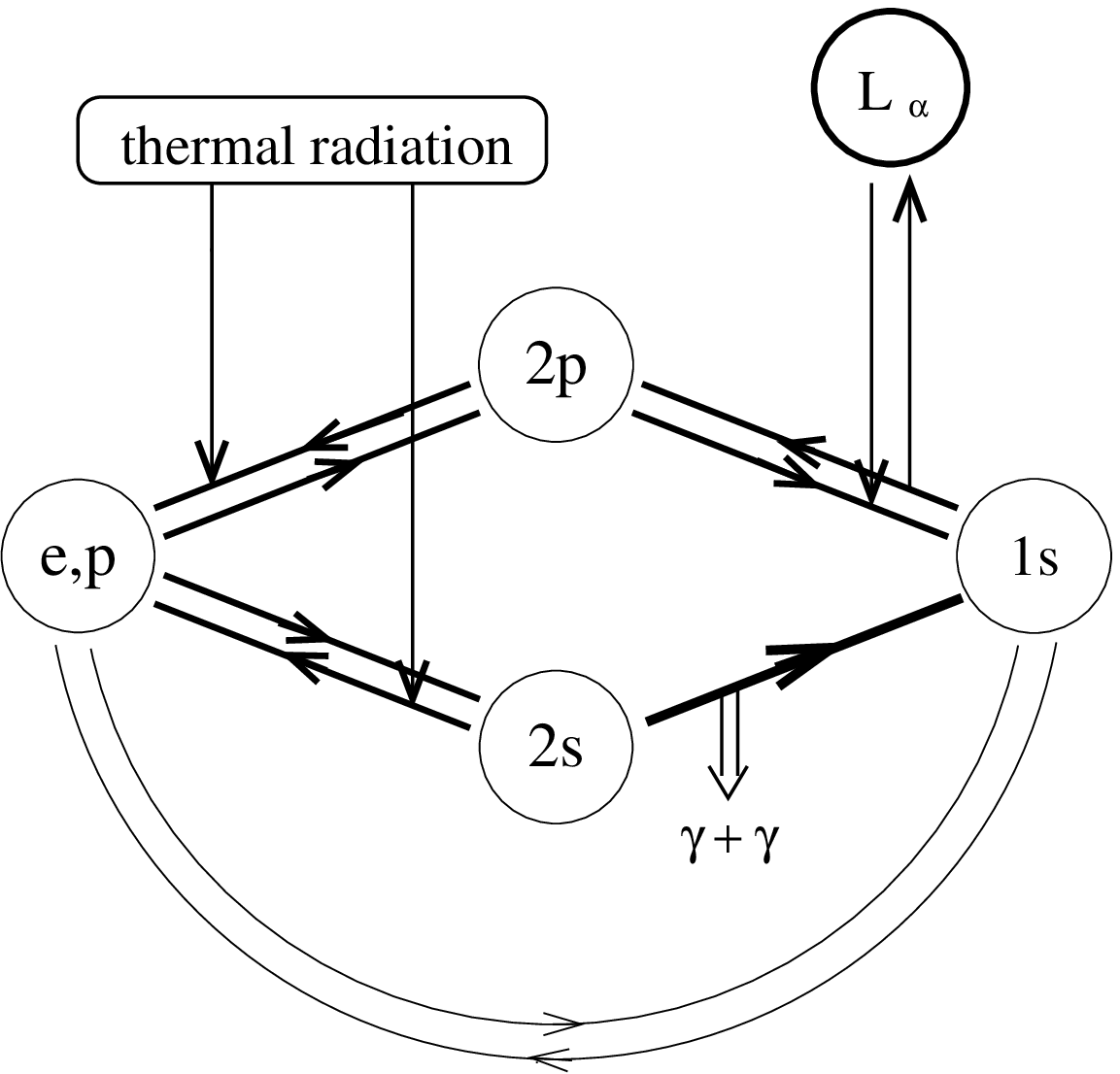} \vspace{0.1cm}\\[0pt]
figure 2
\end{center}
\end{figure}

Second, the thermal radiation is still very efficient and plays the dominant
role in the ionization of the excited hydrogen atom $2s,p-$states (at least,
at the beginning of recombination). Actually to ionize the excited hydrogen
atom the energy of the photon should be only one quarter of the binding
energy $B_{H}.$ The number of such photons is still bigger that the number
of highly energetic photons emitted in recombination acts and therefore
considering the ionization of the excited atoms one can safely ignore the
distortions of the thermal radiation spectrum. On the contrary, these
thermal quanta do not play any essential role in the transitions between $1s$
and $2p,s-$states. The transitions $1s\rightarrow 2p$ are mostly due to $%
L_{\alpha }-$quanta which at the beginning are present in the same number as
the neutral atoms in the ground state. After the degree of ionization
significantly drops the free electrons and the excited states are
overabundant compared to what one would expect according to the equilibrium
Saha's formula. This is why we can neglect $1s+\gamma +\gamma \rightarrow 2s$
transitions compared to the two-photon decay of $2s-$state: $2s\rightarrow
1s+\gamma +\gamma .$ The probability of this process $\left(
W_{2s\rightarrow 1s}\simeq 8.23\mbox{ }sec^{-1}\right) $ is much smaller
than the probability of $2p\rightarrow 1s+L_{\alpha }-$decay $\left(
W_{2s\rightarrow 1s}\simeq 4\times 10^{8}\mbox{ }sec^{-1}\right) .$
Nevertheless, it plays the main role in the nonequilibrium recombination
being, in fact, responsible for the net change of the concentrations of all
\textquotedblleft elements".

The $L_{\alpha }$ quanta emitted in $2p\rightarrow 1s$ transitions are fast
reabsorbed by the hydrogen atoms in the ground state, and these atoms go
back to the \textquotedblleft $2p-$reservoir". Therefore, the main source of
the irreversible \textquotedblright leakage\textquotedblright\ from
\textquotedblright $e,p-$ to $1s-$reservoir\textquotedblright\ is the two
quanta decay via $2s-$levels and the net change of the electron
concentration is mainly due to this process. All other processes return the
\textquotedblright escaped\textquotedblright\ electrons very fast back to
\textquotedblright $e,p-$reservoir\textquotedblright . Hence, the rate of
the overall decrease of the electron concentration (which is equal to the
increase of the neutral atoms in the ground state) due to the two-photon
decay of $2s-$states is:%
\begin{equation}
\frac{dX_{e}}{dt}=-\frac{dX_{1s}}{dt}=-W_{2s}X_{2s},  \label{Hrate}
\end{equation}%
where the relative concentrations $X_{e}\equiv n_{e}/n_{t},$ $X_{2s}\equiv
n_{2s}/n_{t}$ have been introduced$;$ here $n_{t}$ is the total number
density of all neutral atoms plus electrons. I would like to stress once
more that the equation (\ref{Hrate}) ignores all other irreversable
processes, besides of $2s\rightarrow 1s+\gamma +\gamma $ decay, which could
lead as a final outcome to the neutral hydrogen atoms in the ground state .
As we will see later this assumption is valid until the degree of the
ionization drops to rather small values. After that, at the end of
recombination, when some other irreversable processes (in addition to two
quanta decay) become important, I will correct the main equations to account
for them.

To express $X_{2s}$ through $X_{e}$ let us use the quasi-equilibrium
condition for \textquotedblright $2s-$reservoir\textquotedblright . The
rates of the reactions depicted in the Fig. 2 are very high compared to the
rate of the expansion. Therefore the concentrations of the elements in the
\textquotedblright intermediate reservoirs\textquotedblright\ quickly adjust
their quasi-equilibrium values which are determined by condition that the
\textquotedblright net flux\textquotedblright\ for an appropriate
\textquotedblright reservoir\textquotedblright\ should be equal to zero. For
\textquotedblright $2s-$reservoir\textquotedblright\ this condition takes
the following form:%
\begin{equation}
\left\langle \sigma v\right\rangle _{ep\rightarrow \gamma
2s}n_{e}n_{p}-\left\langle \sigma \right\rangle _{\gamma 2s\rightarrow
ep}n_{\gamma }^{eq}n_{2s}-W_{2s\rightarrow 1s}n_{2s}=0,  \label{Hqeq1}
\end{equation}%
where $\left\langle \sigma v\right\rangle $ are the effective rates of the
appropriate reactions and $n_{\gamma }^{eq}$ is the number density of the
thermal photons. The relation between the cross-sections of the direct and
inverse reactions can be easily found if one notes that in the state of
equilibrium these reactions should compensate each other; hence%
\begin{equation}
\frac{\left\langle \sigma \right\rangle _{\gamma 2s\rightarrow ep}n_{\gamma
}^{eq}}{\left\langle \sigma v\right\rangle _{ep\rightarrow \gamma 2s}}=\frac{%
n_{e}^{eq}n_{p}^{eq}}{n_{2s}^{eq}}=\left( \frac{Tm_{e}}{2\pi }\right)
^{3/2}\exp \left( -\frac{B_{H}}{4T}\right)  \label{Hreacratio}
\end{equation}%
where in the second equality I used the Saha's formula and took into account
that the binding energy of $2s-$state is $B_{H}/4.$ Using this relation we
can express $X_{2s}$ from (\ref{Hqeq1}) as%
\begin{equation}
X_{2s}=\left( \frac{W_{2s}}{\left\langle \sigma v\right\rangle
_{ep\rightarrow 2s}}+\left( \frac{Tm_{e}}{2\pi }\right) ^{3/2}\exp \left( -%
\frac{B_{H}}{4T}\right) \right) ^{-1}n_{t}X_{e}^{2}  \label{HXs}
\end{equation}%
Substituting this expression into (\ref{Hrate}) we obtain%
\begin{equation}
\frac{dX_{e}}{dt}=-W_{2s}\left( \frac{W_{2s}}{\left\langle \sigma
v\right\rangle _{ep\rightarrow 2s}}+\left( \frac{Tm_{e}}{2\pi }\right)
^{3/2}\exp \left( -\frac{B_{H}}{4T}\right) \right) ^{-1}n_{t}X_{e}^{2}
\label{HXs1}
\end{equation}%
When the first term inside the bracket is small compared to the second one
the electron and \textit{excited} \textit{states} of hydrogen atoms are in
equilibrium with each other and with thermal radiation. In this case the
last term in the equation (\ref{Hqeq1}) is small compared to the other terms
and the relative concentrations of $e,p$ and $2s-$states still satisfy the
appropriate Saha's relation (r.h.s. equality in (\ref{Hreacratio})). Of
course, it does not mean that the ionization degree in this case is given by
the equilibrium Saha's formula, which is derived under assumption that $1s-$%
state is also in thermal equilibrium with the other states. As I mentioned,
the ground state drops out of equilibrium with the other levels soon after
recombination begins and there is an overabundance of the atoms in the
excited states compared to what one would expect according to the
equilibrium Saha's formula\footnote{%
At the beginning of recombination the rate of change of the hydrogen atoms
in $1s-$state is proportional to $W_{2s\rightarrow 1s}n_{2s}.$This rate is
much smaller than the rate of the reactions $ep\leftrightarrows \gamma 2s,$
determining the concentration of $2s-$states.}.

\bigskip The rate of the recombination to $2s-$level is well approximated by
the formula (see, for instance, \cite{peebles}):%
\begin{equation}
\left\langle \sigma v\right\rangle _{ep\rightarrow \gamma 2s}\simeq
6.3\times 10^{-14}\left( \frac{B_{H}}{4T}\right) ^{1/2}\mbox{ }\frac{cm^{3}}{%
sec}  \label{Hcrsect}
\end{equation}%
and one can easily verify that two terms inside the brackets in (\ref{HXs})
becomes comparable at the temperature $\simeq 2450%
{{}^\circ}%
K.$ Hence only at the temperatures higher than $2450%
{{}^\circ}%
K$ the $e-p$ recombination processes are faster than the two photon decay
and thermal radiation is efficient in keeping the chemical equilibrium
between $e,p$ and $2s-$states.

As long as the temperature drops below this value the photoionization of $2s-
$states becomes less efficient than their two quantum decay. The thermal
radiation does not play essential role after that and the quasi-equilibrium
concentration of $2s-$ states is regulated by the balance of the
recombination rate to $2s-$levels and their two quanta decay rate (the
second term in the equation (\ref{Hqeq1}) can be neglected compared to the
third one). In this case the second term inside the brackets in (\ref{HXs1})
is small compared to the first one and the rate of recombination due to the
leakage of the electron from \textquotedblleft $e,p-$reservoir" through $2s$
reservoir is proportional to $\left\langle \sigma v\right\rangle
_{ep\rightarrow \gamma 2s}n_{t}X_{e}^{2}$ and does not depend on $W_{2s}$ $.$
It is entirely determined by the rate of the recombination to $2s$ level. At
the same time $2p-$ states also drop out equilibrium with electrons, protons
and thermal radiation and most of $L_{\alpha }$ are destroyed in two quanta
decays. As a result the \textquotedblright $e,p\rightarrow 2p\rightarrow 1s-$%
channel\textquotedblright\ becomes also efficient in converting the free
electrons and protons into the neutral hydrogen and increases the
\textquotedblright leakage of the electrons from $e,p-$reservoir%
\textquotedblright . Moreover, nearly every recombination act in one of the
excited states lead to the formation of the neutral hydrogen atom. This
effect is relevant only at the late stages of recombination and can be
easily taken into account if we substitute in (\ref{HXs1}) instead of $%
\left\langle \sigma v\right\rangle _{ep\rightarrow \gamma 2s}$ the rate for
recombination to all \textit{excited} states, which is well approximated by
the fitting formula (see, for instance, \cite{peebles}) 
\begin{equation}
\left\langle \sigma v\right\rangle _{rec}\simeq 8.7\times 10^{-14}\left( 
\frac{B_{H}}{4T}\right) ^{0.8}\mbox{ }\frac{cm^{3}}{sec}  \label{recextot}
\end{equation}%
It is convenient to rewrite the equation (\ref{HXs1}) using instead of
temperature and cosmological time, related to the temperature via (\ref{ct}%
), the redshift parameter $z+1=T/T_{\gamma 0}$. Substituting the numerical
values for the reaction rates and the number density $n_{t}$ in the obtained
equation after elementary calculations we get:%
\begin{equation}
\frac{dX_{e}}{dz}=15.\,\allowbreak 3\frac{\Omega _{b}h_{75}}{\sqrt{\Omega
_{m}}}\left( 0.72\left( \frac{z}{14400}\right) ^{0.3}+10^{4}z\exp \left( -%
\frac{14400}{z}\right) \right) ^{-1}X_{e}^{2},  \label{Heq}
\end{equation}%
when we neglect unity compared to $z.$ This equation can be easily
integrated:%
\begin{equation}
X_{e}\left( z\right) =6.53\times 10^{-2}\frac{\sqrt{\Omega _{m}}}{\Omega
_{b}h_{75}}\left( \int_{z}\frac{dz}{\left( 0.72\left( z/\left( 1.44\times
10^{4}\right) \right) ^{0.3}+10^{4}z\exp \left( -1.44\times 10^{4}/z\right)
\right) }\right) ^{-1}  \label{Hsol1}
\end{equation}%
One can verify that the solution $X_{e}\left( z\right) $ is not very
sensitive to the \textquotedblleft initial conditions" which could be taken
at $z_{in}>z$ (for instance, at $T\simeq 3500%
{{}^\circ}%
K$), when $X_{e}\left( z_{in}\right) \gg X_{e}\left( z\right) .$ The main
contribution to the integral in this case give $z<z_{in}$. At $z>900$
(appropriately at the temperature $T>2450%
{{}^\circ}%
K$) the first term inside the bracket in the integrand can be neglected. In
this case, the expression (\ref{Hsol1}) is well approximated by the formula%
\cite{SZ}:%
\begin{equation}
X_{e}\left( z\right) \simeq \allowbreak 9.\,\allowbreak 1\times 10^{6}\frac{%
\sqrt{\Omega _{m}}}{\Omega _{b}h_{75}}z^{-1}\exp \left( -\frac{14400}{z}%
\right)   \label{Hsol2}
\end{equation}%
and the overall rate of the recombination is completely determined by the
rate of two quanta decay. It is clear from the derivation of (\ref{Hsol1})
that this formula and, correspondingly (\ref{Hsol2}), are applicable only
when the degree of ionization drops significantly below unity and the
deviations from the full equilibrium become quite essential. As a rough
criteria for the applicability of these formulae let us take the moment when
the concentration of the neutral hydrogen reaches about ten percent.
According to (\ref{Hsol2}) for the realistic values of the cosmological
parameters: $\Omega _{m}h_{75}^{2}\simeq 0.3$ and $\Omega
_{b}h_{75}^{2}\simeq 0.03$ this happens at $z\sim 1220$ (appropriately $%
T\sim 3300\div 3400%
{{}^\circ}%
K$) $.$ Therefore in this case the range of applicability of (\ref{Hsol2})
is not very big, namely, $1200>z>900.$ However during this time when the
temperature drops only from $3400%
{{}^\circ}%
K$ to $2450%
{{}^\circ}%
K$ the degree of ionization \ decrease very substantially; at $T\simeq 2450%
{{}^\circ}%
K$ it constitutes $X_{e}\left( 900\right) \simeq 2\times 10^{-2}.$ It is
interesting to compare this result to the prediction of the equilibrium
Saha's formula. According to the equilibrium Saha's formula $X_{e}(3400%
{{}^\circ}%
K)\sim 10^{-1}$ and $X_{e}(2450%
{{}^\circ}%
K)\sim 10^{-5},$ that is, at $z\simeq 900$ the ionization degree exceed the
equilibrium one more than in thousand times. Hence, the deviation from the
equilibrium very essentially delay the recombination process. The other
interesting thing is that the equilibrium ionization degree depends only on
the baryon number density, while in (\ref{Hsol1}) enters also the density of
the cold matter. It is not surprising, since the cold matter determines the
rate of the cosmological expansion which is very important for kinetics when
the deviations from equilibrium become essential.

When the temperature drops below $2450%
{{}^\circ}%
K$ at $z<900$ the approximate formula (\ref{Hsol2}) is not valid anymore and
we have to use (\ref{Hsol1}). The degree of ionization first continues to
drop and finally freezes-out; for instance, for $\Omega _{m}h_{75}^{2}\simeq
0.3$ and $\Omega _{b}h_{75}^{2}\simeq 0.03$ the formula (\ref{Hsol1}) gives: 
$X_{e}\left( z=800\right) \simeq 5\times 10^{-3},$ $X_{e}\left( 400\right)
\simeq 7\times 10^{-4}$ and $X_{e}\left( 100\right) \simeq 4\times 10^{-4}.$
To calculate the freeze-out concentration we note that the integral in (\ref%
{Hsol1}) converges for $z=0$ and is about $4\times 10^{3};$ hence%
\begin{equation}
X_{e}^{f}\simeq 1.6\times 10^{-5}\frac{\sqrt{\Omega _{m}}}{\Omega _{b}h_{75}}
\label{Hfr}
\end{equation}%
After ionization degree drops below unity the approximate results given (\ref%
{Hsol1}) and (\ref{Hsol2}) are in very good agreement with the numerical
solutions of the kinetic equations, while the Saha's approximation do not
reproduce the ionization behavior even roughly.

At the beginning of recombination most of the neutral hydrogen atoms were
formed as a result of the cascade transitions and the number of $L_{\alpha }-
$photons was about the same as a number of hydrogen atoms. What happens with
all these $L_{\alpha }-$photons afterwards? Will they survive and, if so,
could we observe them today as an appropriately redshifted narrow line in
the spectrum of CMB? During the recombination the number density of the $%
L_{\alpha }-$quanta $n_{\alpha }$ is determined by the quasi-equilibrium
condition for \textquotedblright $L_{\alpha }-$reservoir\textquotedblright 
\begin{equation}
W_{2p\rightarrow 1s}n_{2p}=\left\langle \sigma _{\alpha }\right\rangle
n_{\alpha }n_{1s}.  \label{Hqela}
\end{equation}%
Since $n_{1s}\approx n_{t}$ and $n_{2p}\propto X_{e}^{2}$ we see that the
number of these quanta drops proportionally to the ionization degree
squared. Thus, nearly all $L_{\alpha }-$photons which emerged at the
beginning disappear because they are \textquotedblright de
facto\textquotedblright\ destroyed due to the two-photons decay of $2s-$%
states. Therefore there will be no sharp line in the primordial radiation
spectrum. Nevertheless as a result of recombination this spectrum will be
significantly warped in the Wien region. Unfortunately, the spectrum
distortions lie in those part of the spectrum, where they are strongly
saturated by the radiation from the other astrophysical sources and one can
not observationally verify this important consequence of the hydrogen
recombination.

Finally let us find when the universe becomes transparent for the radiation.
It happens when the typical time between the photon scattering begins to
exceed the cosmological time. The Raleigh's cross-section for the scattering
on the neutral hydrogen is negligibly small and in spite of their low
concentration, the main role in opaqueness play the free electrons . The
cross-section of the scattering on free electron is equal to $\sigma
_{T}\simeq 6.65\times 10^{-25}$ $cm^{2}$ and the equation defining the
moment when the radiation completely decouples form the matter takes the
form:%
\begin{equation}
\frac{1}{\sigma _{T}n_{t}X_{e}}\sim t_{cosm},  \label{Hrd}
\end{equation}%
This equation can be rewritten as%
\begin{equation}
X_{e}^{dec}\sim 40\frac{\sqrt{\Omega _{m}}}{\Omega _{b}h_{75}}\left( \frac{%
T_{\gamma 0}}{T_{dec}}\right) ^{3/2}  \label{Hxdec}
\end{equation}%
By \textquotedblright try-out\textquotedblright\ one can easily check that
the decoupling happens at $T_{dec}\sim 2500%
{{}^\circ}%
K$ ( the corresponding redshift $z_{dec}\sim 900$) irrespective how big are
the values of the cosmological parameters. If $\Omega _{m}h_{75}^{2}\simeq
0.3$ and $\Omega _{b}h_{75}^{2}\simeq 0.03$ the ionization degree at this
moment is about\footnote{%
It is rather interesting to note that this time coincides with the moment
when $e,p$ and $2s-$levels come out of equilibrium and the approximate
formula (\ref{Hsol2}) becomes inapplicable.} $2\times 10^{-2}$.

\section{Asymptotic behavior of the transfer functions}

The resulting fluctuations of the background radiation depend on the
gravitational potential $\Phi $ and the radiation energy density
fluctuations $\delta _{\gamma }\equiv \delta \varepsilon _{\gamma
}/\varepsilon _{\gamma }$ at the moment of recombination. To determine these
quantities we have to study the gravitational instability in two component
medium consisting of the coupled baryon-radiation plasma and the cold dark
matter. Because these components interact only gravitationally their
energy-momentum tensors conserve separately. In the cosmological conditions
the shear viscosity can not be neglected for the baryon-radiation plasma and
leads to the dissipation of perturbations in small scales (Silk damping).
For imperfect fluid with the energy density $\varepsilon $ and the pressure $%
p$ one can use the energy-momentum tensor given in\footnote{%
I will neglect the heat conduction since it does not change substantially
the Silk damping scale.} \cite{W}. Then one can easily find that in a
homogeneous universe with small perturbations described by the metric (\ref%
{metric}) the conservation laws $T_{\beta ;\alpha }^{\alpha }=0$ in the
first order in perturbations reduce to 
\begin{equation}
\delta \varepsilon ^{\prime }+3\mathcal{H}\left( \delta \varepsilon +\delta
p\right) -3\left( \varepsilon +p\right) \Phi ^{\prime }+a\left( \varepsilon
+p\right) u_{\ \ ,i}^{i}=0.  \label{R1}
\end{equation}%
\begin{equation}
\frac{1}{a^{4}}\left( a^{5}\left( \varepsilon +p\right) u_{\ \
,i}^{i}\right) ^{\prime }-\frac{4}{3}\mathbf{\eta }\Delta u_{\ \
,i}^{i}+\Delta \delta p+\left( \varepsilon +p\right) \Delta \Phi =0.
\label{R2}
\end{equation}%
where $\delta \varepsilon ,\delta p$ are, respectively, the perturbations of
the energy density and pressure ; $u^{i}$ is the peculiar 3-velocity and $%
\mathbf{\eta }$ is the shear viscosity coefficient. Note that the first
equations which follows from $T_{0;\alpha }^{\alpha }=0$ does not contain
the shear viscosity. The second equation was obtained by taking the
divergence of the equations $T_{i;\alpha }^{\alpha }=0.$ As it was already
noted, these two equations are \textit{separately} valid for the dark matter
and for the baryon-radiation plasma.

\textit{Dark matter.} For dark matter, the pressure $p$ and the shear
viscosity $\mathbf{\eta }$ are both equal to zero. Taking into account that $%
\varepsilon _{d}a^{3}=const$ we obtain from (\ref{R1}) that the fractional
perturbations in the energy of dark matter component $\delta _{d}\equiv
\delta \varepsilon _{d}/\varepsilon _{d}$ satisfy the equation%
\begin{equation}
\left( \delta _{d}-3\Phi \right) ^{\prime }+au_{\mbox{ },i}^{i}=0.
\label{R3}
\end{equation}%
If we express $u_{,i}^{i}$ in terms of $\delta _{d}$ and $\Phi $ and
substitute it into (\ref{R2}) then the resulting equation takes the
following form: 
\begin{equation}
\left( a\left( \delta _{d}-3\Phi \right) ^{\prime }\right) ^{\prime
}-a\Delta \Phi =0.  \label{R4}
\end{equation}

\textit{Radiation-baryon plasma}. The baryons and radiation are tightly
coupled before recombination and, therefore, generically only the sum of
their energy-momentum tensors satisfies the conservation laws (\ref{R1}) and
(\ref{R2}). Nevertheless, in particular case when the baryons are
nonrelativistic, the equation (\ref{R1}) is still valid separately by
baryons and radiation, because the energy conservation law for the baryons, $%
T_{0;\alpha }^{\alpha }=0,$ reduces in this case to the conservation law for
the total baryon number. (Of course this is not true for (\ref{R2})) since
baryons and radiation \textquotedblleft move together" and there exist
momentum exchange between these components.) Hence, the fractional density
fluctuations in baryons, $\delta _{b}\equiv \delta \varepsilon
_{b}/\varepsilon _{b},$ satisfy the equation similar to (\ref{R3}):%
\begin{equation}
\left( \delta _{b}-3\Phi \right) ^{\prime }+au_{\mbox{ },i}^{i}=0.
\label{R4a}
\end{equation}%
As it follows from (\ref{R1}) the corresponding equation for the
perturbations in the radiation component, $\delta _{\gamma }\equiv \delta
\varepsilon _{\gamma }/\varepsilon _{\gamma }$ takes the form 
\begin{equation}
\left( \delta _{\gamma }-4\Phi \right) ^{\prime }+\frac{4}{3}au_{\mbox{ }%
,i}^{i}=0.  \label{R4b}
\end{equation}%
Since the photons and baryons are tightly coupled their velocities are the
same. Therefore multiplying (\ref{R4b}) by $3/4$ and subtracting equation (%
\ref{R4a}) we obtain 
\begin{equation}
\frac{\delta s}{s}\equiv \frac{3}{4}\delta _{\gamma }-\delta _{b}=const
\label{R4c}
\end{equation}%
where $\delta s/s$ are the fractional entropy fluctuations in the
baryon-radiation plasma. For adiabatic perturbations, $\delta s=0$ and,
therefore, we have 
\begin{equation}
\delta _{b}=\frac{3}{4}\delta _{\gamma }.  \label{R5}
\end{equation}%
If we express $u_{,i}^{i}$ in terms of $\delta _{\gamma }$ and $\Phi $ from (%
\ref{R4b}) and substitute into (\ref{R2}), we obtain 
\begin{equation}
\left( \frac{\delta _{\gamma }^{\prime }}{c_{s}^{2}}\right) ^{\prime }-\frac{%
3\mathbf{\eta }}{\varepsilon _{\gamma }a}\Delta \delta _{\gamma }^{\prime
}-\Delta \delta _{\gamma }=\frac{4}{3c_{s}^{2}}\Delta \Phi +\left( \frac{%
4\Phi ^{\prime }}{c_{s}^{2}}\right) ^{\prime }-\frac{12\mathbf{\eta }}{%
\varepsilon _{\gamma }a}\Delta \Phi ^{\prime },  \label{R5a}
\end{equation}%
where $\Delta $ is the Laplacian and $c_{s}^{2}$ is the squared speed of
sound in the baryon-radiation plasma, which is equal to: 
\begin{equation}
c_{s}^{2}\equiv \frac{\delta p}{\delta \varepsilon }=\frac{\delta p_{\gamma }%
}{\delta \varepsilon _{\gamma }+\delta \varepsilon _{b}}=\frac{1}{3}\left( 1+%
\frac{3}{4}\frac{\varepsilon _{b}}{\varepsilon _{\gamma }}\right) ^{-1}.
\label{R6}
\end{equation}%
Without taking into account the polarization effects the shear viscosity
coefficient entering (\ref{R5a}) is given by\cite{W}: 
\begin{equation}
\mathbf{\eta }=\frac{4}{15}\varepsilon _{\gamma }\tau _{\gamma }  \label{R6b}
\end{equation}%
where $\tau _{\gamma }$ is the mean free time for the photons.

Thus we derived two perturbation equations (\ref{R4}) and (\ref{R5a}), which
being supplemented by 0-0 component of the Einstein equations \cite{MB} 
\begin{equation}
\Delta \Phi -3\mathcal{H}\Phi ^{\prime }-3\mathcal{H}^{2}\Phi =4\pi
Ga^{2}\left( \varepsilon _{d}\delta _{d}+\frac{1}{3c_{s}^{2}}\varepsilon
_{\gamma }\delta _{\gamma }\right)   \label{R10}
\end{equation}%
form a closed system of equations for three unknown variables $\delta
_{d},\delta _{\gamma }$ and $\Phi $ (we used (\ref{R5}) to express $\delta
_{b}$ in terms of $\delta _{\gamma }$)$.$

>From (\ref{R4b}) it follows the useful relation for \textit{only the
radiation contribution} to the divergence of $0-i$ components of the
energy-momentum tensor, 
\begin{equation}
T_{0,i}^{i}=\frac{4}{3}\varepsilon _{\gamma }u_{0}u_{,i}^{i}=\left( 4\Phi
-\delta _{\gamma }\right) ^{\prime }\varepsilon _{\gamma }\,\,\,
\label{R10a}
\end{equation}%
which is used in (\ref{F0im})

\textit{Longwave perturbations: }$\left( k\ll \eta _{r}^{-1}\right) $ The
behavior of perturbations strongly depends on how big is their scales
compared to the horizon. First I consider the long wavelength perturbations
with $k\eta _{r}\ll 1$ ($k$ is comoving wavenumber) which cross the horizon
only after recombination. Knowing the gravitational potential we can easily
find $\delta _{\gamma }$. In fact for this longwave perturbations one can
neglect the velocity term in the equation (\ref{R4b}), which after that can
be easily integrated with the result 
\begin{equation}
\delta _{\gamma }-4\Phi =C,  \label{R11}
\end{equation}%
where $C$ is the constant of integration. To determine $C$, we note that,
during the radiation dominated epoch, the gravitational potential is mostly
due to the fluctuations in the radiation component and does not change on
supercurvature scales. At early times, $\delta _{\gamma }\simeq -2\Phi
\left( \eta \ll \eta _{eq}\right) \equiv -2\Phi ^{0}$ (see \cite{MB}) ;
hence $C=-6\Phi ^{0}.$ After equality, when the dark matter overtakes the
radiation, the gravitational potential $\Phi $ changes its value by factor
of $9/10$ and then remains constant, that is, $\Phi \left( \eta \gg \eta
_{eq}\right) =\left( 9/10\right) \Phi ^{0}.$ Therefore, if cold dark matter
dominates at recombination, it follows from (\ref{R11}) that 
\begin{equation}
\delta _{\gamma }\left( \eta _{r}\right) =-6\Phi ^{0}+4\Phi \left( \eta
_{r}\right) =-\frac{8}{3}\Phi \left( \eta _{r}\right) .  \label{R12}
\end{equation}%
One arrives at the same conclusion by noting that, for the adiabatic
perturbations, $\delta _{\gamma }=4\delta _{d}/3$ and $\delta _{d}\simeq
-2\Phi \left( \eta _{r}\right) $ at recombination.

\textit{Intermediate scales }$(\eta _{r}^{-1}<k<\eta _{eq}^{-1})$ Next I
consider the scales which enter horizon in between the equality $\left( \eta
_{eq}\right) $ and recombination $\left( \eta _{r}\right) $. The
perturbations which enter horizon within this rather short time interval are
especially interesting since they are responsible for the first few acoustic
peaks in the CMB spectrum. Unfortunately, for the realistic values of the
cosmological parameters the solution for these perturbations cannot be found
analytically with needed accuracy because in the realistic models the
condition $\eta _{eq}\ll \eta _{r}$ is not satisfied. Nevertheless to gain
an intuition about the behavior of perturbations, it is very useful to
consider the models where $\eta _{eq}\ll \eta _{r}$ and derive the
appropriate asymptotic expressions for the perturbations with $\eta
_{r}^{-1}\ll k\ll \eta _{eq}^{-1}.$ To simplify the consideration I also
assume that the contribution of baryons to the gravitational potential is
negligible compared to the contribution of the cold dark matter.

In general, there exist four instability modes in two component medium. The
set of equations is rather complicated and they can not be solved
analytically without making further assumptions. However in our case, the
problem can be simplified if we note that if the perturbation enters horizon
sufficiently late after equality ($\eta \gg \eta _{eq}$), the appropriate
gravitational potential, which is mainly due to the perturbations in the
cold dark matter component, remains unchanged and stays constant afterwards $%
\left( \Phi _{k}\left( \eta \right) =const\right) $ \cite{MB}. The baryons
do not contribute much to the gravitational potential, however they can
still significantly influence the speed of sound after equality.

Under assumption we have made, the gravitational potential $\Phi $ can be
considered as an external source in equation (\ref{R5a}). Therefore, the
general solution of this equation is given by the sum of a general solution
of homogeneous equation (with $\Phi =0$) and a particular solution of (\ref%
{R5a}). Introducing the variable $x,$ defined by $dx=c_{s}^{2}d\eta ,$ and
taking into account that the time derivatives of the potential (\ref{R5a})
are equal to zero ($\Phi =const$), we reduce the equation (\ref{R5a}) to 
\begin{equation}
\frac{d^{2}\delta _{\gamma }}{dx^{2}}-\frac{4\tau _{\gamma }}{5a}\Delta 
\frac{d\delta _{\gamma }}{dx}-\frac{1}{c_{s}^{2}}\Delta \delta _{\gamma }=%
\frac{4}{3c_{s}^{4}}\Delta \Phi .  \label{R10ab}
\end{equation}%
where the second term is due to the viscosity. If the speed of sound is
slowly varying, this equation has an obvious approximate solution 
\begin{equation}
\delta _{\gamma }\simeq -\frac{4}{3c_{s}^{2}}\Phi .  \label{R11ab}
\end{equation}%
The general solution of the homogeneous equation (\ref{R10ab}) can be
obtained in the WKB approximation. Let us consider the plane wave
perturbation with the comoving wavenumber $k.$ Introducing instead of $%
\delta _{\gamma }$ the new variable 
\begin{equation}
y\equiv \delta _{\gamma }\exp \left( \frac{2}{5}k^{2}\int \frac{\tau
_{\gamma }}{a}dx\right) ,  \label{Rdef}
\end{equation}%
we find from (\ref{R10ab}) that it satisfies the equation%
\begin{equation}
\frac{d^{2}y}{dx^{2}}+\frac{k^{2}}{c_{s}^{2}}\left( 1-\frac{4c_{s}^{2}}{25}%
\left( \frac{k\tau _{\gamma }}{a}\right) ^{2}-\frac{2c_{s}^{2}}{5}\left( 
\frac{\tau _{\gamma }}{a}\right) ^{\prime }\right) y=0.  \label{Reqy}
\end{equation}%
For the perturbations with the scale ($\lambda _{ph}\sim a/k$) bigger than
the mean free path of the photons\footnote{%
In fact, the imperfect fluid approximation can be used only in this case} ($%
\sim \tau _{\gamma }$)$,$ the second term inside the brackets is negligible.
The third term which is about $\tau _{\gamma }/a\eta \sim \tau _{\gamma }/t$ 
$\ll 1$ can be also skipped. Therefore the WKB solution for $y$ is 
\begin{equation}
y\simeq \sqrt{c_{s}}\left( C_{1}\cos \left( k\int \frac{dx}{c_{s}}\right)
+C_{2}\sin \left( k\int \frac{dx}{c_{s}}\right) \right) .  \label{Rsoly}
\end{equation}%
Returning back to $\delta _{\gamma }$ $($see (\ref{Rdef})) and combining
this solution with (\ref{R11ab}), we obtain 
\begin{equation}
\delta _{\gamma }\simeq -\frac{4}{3c_{s}^{2}}\Phi _{k}+\sqrt{c_{s}}\left(
C_{1}\cos \left( k\int c_{s}d\eta \right) +C_{2}\sin \left( k\int c_{s}d\eta
\right) \right) e^{-\left( k/k_{D}\right) ^{2}}.  \label{Rsolfin}
\end{equation}%
Here we have introduced the dissipation scale characterized by the comoving
wavenumber:%
\begin{equation}
k_{D}\left( \eta \right) \equiv \left( \frac{2}{5}\int_{0}^{\eta }c_{s}^{2}%
\frac{\tau _{\gamma }}{a}d\eta \right) ^{-1/2}.  \label{Rdiss}
\end{equation}%
In the limit of constant speed of sound and vanishing viscosity the solution
(\ref{Rsolfin}) is exact and valid also in the limit $k\rightarrow 0.$

>From (\ref{Rsolfin}), it is clear that the viscosity efficiently damps the
perturbations on comoving scales $\lambda \leq 1/k_{D}$. Using the formula (%
\ref{Rdiss}) with $c_{s}^{2}=1/3$ and, assuming instantaneous recombination
we obtain the following estimate for the dissipation scale 
\begin{equation}
\left( k_{D}\eta _{r}\right) ^{-1}\simeq 0.6\left( \Omega _{m}h^{2}\right)
^{1/4}\left( \Omega _{b}h^{2}\right) ^{-1/2}z_{r}^{-3/4}.  \label{Rd1}
\end{equation}

The constants of integration $\ C_{1}$ and $C_{2}$ in (\ref{Rsolfin}) can be
determined if we note that at earlier stages when the speed of sound does
not change too much the solution (\ref{Rsolfin}) is also valid when the
scale of perturbation still exceeds the horizon scale. As we have found
before the amplitude of the longwave perturbations ($k\eta \ll 1$) is equal
to $\delta _{\gamma }\simeq -8\Phi _{k}/3=const$ at $\eta \gg \eta _{eq}.$
Assuming that at the moment when the perturbation enters the horizon the
speed of sound is still not very different from $1/\sqrt{3}$ we find that $%
C_{1}=4/3^{3/4}$ and $C_{2}=0$; hence 
\begin{equation}
\delta _{\gamma }\left( \eta \right) =\left[ -\frac{4}{3c_{s}^{2}}+\frac{4%
\sqrt{c_{s}}}{3^{3/4}}e^{-\left( k/k_{D}\right) ^{2}}\cos \left(
k\int_{0}^{\eta }c_{s}d\eta ^{\prime }\right) \right] \left( \frac{9}{10}%
\Phi _{k}^{0}\right)  \label{R1com}
\end{equation}%
for $k\ll \eta _{eq}^{-1}.$ Here we took into account that $\Phi _{k}=9\Phi
_{k}^{0}/10$ and expressed the result in terms of the initial gravitational
potential on superhorizon scales before equality $\Phi _{k}^{0}.$ The result
(\ref{R1com}) coincides with the result obtained by S. Weinberg \cite{W} in
synchronous coordinate system.

\textit{Shortwave perturbations }$(k\gg \eta _{eq}^{-1})$ Finally I consider
the perturbations which enter the curvature scale before equality. At $\eta
\ll \eta _{eq}$, the radiation dominates and in this case the appropriate
expressions for $\Phi $ and $\delta _{\gamma }$ were derived, for instance,
in \cite{MB}. Neglecting the decaying mode we find that after perturbation
entered the horizon, that is, at $k^{-1}\ll \eta \ll \eta _{eq}$ 
\begin{equation}
\delta _{\gamma }\simeq 6\Phi _{k}^{0}\cos \left( k\eta /\sqrt{3}\right) ,%
\mbox{ \ \ }\Phi _{k}\left( \eta \right) \simeq -\frac{9\Phi _{k}^{0}}{%
\left( k\eta \right) ^{2}}\cos \left( k\eta /\sqrt{3}\right) .  \label{Rdg}
\end{equation}%
The dissipation, which becomes important only before recombination, can be
treated similar to how it was done above. Therefore I neglect the
dissipation term here and restore the damping factors only in the final
expressions.

After inhomogeneity entered the horizon, the cold dark matter starts
\textquotedblleft to slide" with respect to the radiation. To get an idea
about the behavior of the inhomogeneities in the cold dark matter component
itself we can use the equation (\ref{R4}), which after integration becomes 
\begin{equation}
\delta _{d}=3\Phi +\int \frac{d\eta ^{\prime }}{a}\int a\Delta \Phi d\eta
^{\prime \prime }.  \label{Rdb}
\end{equation}%
Note that this is an exact relation which is always valid for any $k.$
During the radiation-dominated epoch, the main contribution to the
gravitational potential is due to the radiation, and, therefore, the
gravitational potential in the equation (\ref{Rdb}) can be treated as an
external source. We can fix the constant of integration in (\ref{Rdb})
substituting there the exact solution for the radiation dominated universe
(see (5.45)-(5.46) in \cite{MB}) and noting that, at earlier times on
superhorizon scales one has to match the well-known result for the longwave
perturbations: $\delta _{d}\simeq 3\delta _{\gamma }/4\simeq -3\Phi
_{k}^{0}/2.$ As a result we obtain that after entering the horizon, but
before equality

\begin{equation}
\delta _{d}\simeq -9\left( \mathbf{C-}\frac{1}{2}+\ln \left( k\eta /\sqrt{3}%
\right) +O\left( \left( k\eta \right) ^{-1}\right) \right) \Phi _{k}^{0},
\label{Rdd}
\end{equation}%
where $\mathbf{C}=0.577...$ is the Euler constant. That is the perturbations
in the cold matter component are ``frozen" (they grow only logarithmically)

>From (\ref{R10}) it is easy to see that before equality the contribution of
dark matter perturbations to the gravitational potential is suppressed by
the factor $\varepsilon _{d}/\varepsilon _{\gamma }$ compared to the
contribution from the radiation component. At equality, the dark matter
begins to dominate and the density perturbation $\delta _{d}$ grow as $%
\propto \eta ^{2}$(see, \cite{MB}). As a result, the appropriate
gravitational potential \textquotedblleft freeze out\textquotedblright\ at
the value 
\begin{equation}
\Phi _{k}\left( \eta >\eta _{eq}\right) \sim \left. -\frac{4\pi
Ga^{2}\varepsilon }{k^{2}}\delta _{d}\right\vert _{\eta _{eq}}\sim O\left(
1\right) \frac{\ln \left( k\eta _{eq}\right) }{\left( k\eta _{eq}\right) ^{2}%
}\Phi _{k}^{0}  \label{R20}
\end{equation}%
and stays constant until the recombination. One can get the exact numerical
coefficients in this formula in the following way. For shortwave
perturbations, the time derivatives of the gravitational potential in (\ref%
{R4}), (\ref{R10}) can be neglected compared to the spatial derivatives.
Then from these equations it follows that%
\begin{equation}
\left( a\delta _{d}^{\prime }\right) ^{\prime }-4\pi Ga^{3}\left(
\varepsilon _{d}\delta _{d}+\frac{1}{3c_{s}^{2}}\varepsilon _{\gamma }\delta
_{\gamma }\right) =0.  \label{R21a}
\end{equation}%
The second term here induces the corrections to the solution (\ref{Rdd})
which become significant only near equality. These corrections are mostly
due to $\varepsilon _{d}\delta _{d}-$term. The term $\propto \varepsilon
_{\gamma }\delta _{\gamma }$ does not lead to essential corrections to the
solution (\ref{Rdd}) before equality and it is also negligible compared to $%
\varepsilon _{d}\delta _{d}-$term after equality; hence it can be skipped in
(\ref{R21a}). As a result the obtained equation can be rewritten in the
following form 
\begin{equation}
x\left( 1+x\right) \frac{d^{2}\delta _{d}}{dx^{2}}+\left( 1+\frac{3}{2}%
x\right) \frac{d\delta _{d}}{dx}-\frac{3}{2}\delta _{d}=0.  \label{Rd2}
\end{equation}%
where $x\equiv a/a_{eq}.$ The general solution of this equation is (see, for
instance, \cite{W})%
\begin{equation}
\delta _{d}=C_{1}\left( 1+\frac{3}{2}x\right) +C_{2}\left[ \left( 1+\frac{3}{%
2}x\right) \ln \frac{\sqrt{1+x}+1}{\sqrt{1+x}-1}-3\sqrt{1+x}\right]
\label{Rd3}
\end{equation}%
At $x\ll 1$ it should coincide with (\ref{Rdd}). Comparing (\ref{Rd3}) with (%
\ref{Rdd}) at $x\ll 1$ we find%
\begin{equation}
C_{1}\simeq -9\left( \ln \left( \frac{2k\eta _{\ast }}{\sqrt{3}}\right) +%
\mathbf{C-}\frac{7}{2}\right) \Phi _{k}^{0},\mbox{ \ \ \ \ \ \ }C_{2}\simeq
9\Phi _{k}^{0}  \label{Rd4}
\end{equation}%
where $\eta _{\ast }=\eta _{eq}/\left( \sqrt{2}-1\right) .$ During the
matter dominated epoch ( $x\gg 1$), the second term in (\ref{Rd3})
corresponds to the decaying mode. Neglecting this mode, assuming that the
baryon contribution to the potential is negligible compared to the dark
matter and using the relation between the gravitational potential and $%
\delta _{d}$ (see (\ref{R10})) one finally gets 
\begin{equation}
\Phi _{k}\left( \eta \gg \eta _{eq}\right) \simeq \frac{\ln \left( 0.15k\eta
_{eq}\right) }{\left( 0.27k\eta _{eq}\right) ^{2}}\Phi _{k}^{0}  \label{Rd6a}
\end{equation}%
in agreement with \cite{W}. The fluctuations in the radiation $\delta
_{\gamma }$ after equality continue to behave as sound waves in the external
gravitational potential given by (\ref{Rd6a}). Therefore, they are described
by (\ref{Rsolfin}), where we have to substitute the potential (\ref{Rd6a})
instead of $\Phi _{k}^{0}$. The constant of integration can be fixed by
comparing the oscillating part of this solution to the result in (\ref{Rdg})
at $\eta \sim \eta _{eq}$. Then, we find that at $\eta \gg \eta _{eq}$ 
\begin{equation}
\delta _{\gamma }\simeq \left[ -\frac{4}{3c_{s}^{2}}\frac{\ln \left(
0.15k\eta _{eq}\right) }{\left( 0.27k\eta _{eq}\right) ^{2}}+3^{5/4}\sqrt{%
4c_{s}}\cos \left( k\int_{0}^{\eta }c_{s}d\eta \right) e^{-\left(
k/k_{D}\right) ^{2}}\right] \Phi _{k}^{0}  \label{Rd7}
\end{equation}%
for $k\gg \eta _{eq}^{-1}$. $\ $\ We have restored here the Silk damping
factor. During the radiation dominated epoch, the damping scale, which is
proportional to the photon mean free path, is very small. However, it
increases just before the recombination and therefore the oscillating
contribution to $\delta _{\gamma }$ is exponentially suppressed on small
scales.

\end{document}